\documentclass[useAMS,usenatbib]{mn2e}
\usepackage{natbib}
\usepackage{graphicx}
\usepackage{amssymb}
\usepackage{amsmath}
\usepackage{mathrsfs}
\usepackage[usenames]{color}
\title{Extreme Gravitational Lensing near Rotating Black Holes}
\author [Kris Beckwith and Chris Done]{Kris Beckwith$^1$ and Chris
Done$^1$ \\ $^1$Department of Physics, University of Durham, South Road,
Durham DH1 3LE, UK}
\date{Released 2004 Xxxxx XX}

\pagerange{\pageref{firstpage}--\pageref{lastpage}} \pubyear{2004}

\def\LaTeX{L\kern-.36em\raise.3ex\hbox{a}\kern-.15em
    T\kern-.1667em\lower.7ex\hbox{E}\kern-.125emX}

\begin{document}

\label{firstpage}

\maketitle

\begin{abstract}
We describe a new approach to calculating photon trajectories and
gravitational lensing effects in the strong gravitational field of the
Kerr black hole. These techniques are applied to explore both the
imaging and spectral properties of photons that perform multiple
orbits of the central mass before escaping to infinity. Viewed at
large inclinations, these higher order photons contribute $\sim 20 \%$
of the total luminosity of the system for a Schwarzschild hole, whilst
for an extreme Kerr black hole this fraction rises to $\sim 60 \%$. In
more realistic models these photons will be re-absorbed by the disc at
large distances from the hole, but this returning radiation could
provide a physical mechanism to resolve the discrepancy between the
predicted and observed optical/UV colours in AGN. Conversely, at low
inclinations, higher order images re-intercept the disc plane close to
the black hole, so need not be absorbed by the disc if this is within
the plunging region. These photons form a bright ring carrying
approximately $10 \%$ of the total disc luminosity for a Schwarzchild
black hole.  The spatial separation between the inner edge of the disc
and the ring is similar to the size of the event horizon. This is
resolvable for supermassive black holes with proposed X-ray
interferometery missions such as MAXIM, so has the potential to
provide an observational test of strong field gravity.

\end{abstract}


\section{Introduction}

Black holes are the ultimate test of strong gravity, spacetime so warped
that not even light can escape. By definition they have no emission (apart
from Hawking radiation), yet their immense gravitational potential energy
can be tapped by any infalling material. This can power a luminous
accretion flow where the emission has its origin close to the black hole
event horizon, as is seen in many objects including Active Galactic
Nuclei, Galactic black hole binaries, Ultra-Luminous X-ray Sources and
Gamma Ray Bursts. Photons emitted in this region are subject to general
relativistic effects such as light-bending, gravitational lensing and
redshift, as well as special relativistic effects as the emitting material
will be moving rapidly (e.g. \citealt{F00}). These are well-understood
from a theoretical standpoint, so accreting objects provide a natural
laboratory to test the properties of strong gravitational fields.

Calculations of the relativistic corrections to photon properties have
been ongoing for nearly three decades, starting with the classic work of
\cite{C75} who calculated the distortions expected on the spectrum of a
geometrically thin, optically thick, Keplerian accretion disc orbiting a
Kerr black hole. Interest in these calculations dramatically increased
with the realisation that the accretion disc could emit {\em line} as
well as continuum radiation. Iron K$\alpha$ fluorescence resulting
from X-ray irradiation of the accretion disc can give a narrow feature, on
which the relativistic distortions are much more easily measured than on
the broad accretion disc continuum \cite{F89}.  Since then, several groups
have developed numerical codes that are capable of determining these
effects both for standard discs \citep{DKY04} and alternative emission
geometries \citep{B04}.

While the problem is well-defined, there are many technical and numerical
issues which arise in calculating the effects of strong gravity.  
Lightbending can result in lensing which strongly amplifies the emission,
so a very small area of the accretion flow can make a large contribution
on the observed flux. Here we describe how to efficiently calculate the
effects of strong lightbending, and illustrate the effectiveness of this
approach by using the code to compute the most extreme lightbending
possible, the higher order images and spectra of an accretion disc \citep{V93,BHO94,C98}.
The first order image is from photons from the underside of the disc which are
bent back into the observers line of sight, while second order images are
produced by photons from the upper side of the disc which complete a full
orbit around the black hole before reaching the observer. Obviously such
paths must cross the equatorial plane, so are likely to re-intercept the
disc. For an optically thick disc then this returning radiation adds to
the intrinsic disc emission, and can enhance the emissivity at small radii
for extreme Kerr black holes though it has little effect for Schwarzchild
\citep{C76,LNP90,AK00}.

Nonetheless, the first order image of the disc can dominate the flux at
high inclinations if the optically thick disc has rather limited radial
extent. Even if it does not, some of the higher order image flux can
escape to the observer through the inner 'hole' in the disc below the
radius of the minimum stable orbit, assuming that the plunging region is
optically thin (but see \citealt{RB97}). The fraction escaping through
this inner 'hole' is rather larger for Schwarzchild than for Kerr, as the
size of the gap between the innermost stable orbit and horizon is larger
for the non-spinning black hole. Obviously, such paths are exquisitely
sensitive to the gravitational potential, being close to the true photon
orbit point which is the (unstable) crossover between capture by the black
hole, and escape to infinity. This makes them potentially an excellent
test of strong gravity, and they could be observable in nearby luminous
AGN with micro-arcsec imaging X-ray interferometers such as MAXIM
(\citealt{G01}).

Such instruments could also observe the {\em spin} of a nearby
supermassive black hole simply from the size of the {\em direct} image
\citep{F03,T04}, assuming that the mass and distance are known. A disc
down to the last stable orbit extends down to 6~$R_g$ in Schwarzchild but
only 1.23~$R_g$ in extreme Kerr. Lightbending is stronger in Kerr than in
Schwarzchild, but the apparent size of the 'hole' in the disc still
changes by a factor of $\sim 3$. This contrasts with the case where the
accretion flow has emission down below the last stable orbit, where the
size of the true black hole shadow is rather similar for both Schwarzchild
and Kerr \citep{FMA00,T04}.  Observations of the galactic black hole
binaries support to the idea that there is a disc down to the minimum
stable orbit in certain, fairly high luminosity spectral states
\citep{E93,K99,GD04}. However, at lower luminosities this is probably
replaced by a more complex accretion flow which may have continuous energy
release down to the horizon \citep{NY95,AK00,KH02,AP03}. While the stellar
remnant black holes require nano-arcsecond imaging to resolve, nearby
luminous AGN also should have a 'hole' in the disc from the minimum stable
orbit which is accessible to MAXIM.

The paper is split into two parts. In Section 2 we introduce the relevant
theory necessary to perform efficient calculation of photon properties in
strong gravity, including a review the properties of the effective
potentials governing photon motion. In Section 3, we apply these
techniques to examine the contribution of orbiting photons to the standard
accretion disc solution through images of the inner region of the
accretion flow, generation of fluorescent Iron K$\alpha$ line profiles and
the relative roles played by the different types of photon paths. Readers
not interested in technical details should go straight to Section 3.

\section{Calculating Photon Paths in Strong Gravity}

\subsection{Null Geodesic Equations}

Material in an accretion flow near to a black hole emits radiation, which
is received by a (possibly distant) observer. These photons enable the
observer to form an image of the flow, which in turn determines the
measured spectral properties \citep{BD04}. In the geometric optics
approximation, the photons follow null geodesics, which in the Kerr metric
(written in Boyer-Lindquist coordinates with $G = M = c =1$) are described
by \citep{C83}: \begin{gather}
   \label{eqn:2.1.2}
   \begin{split}
     p^{r} = \dot{r} = \rho^{-2} \sqrt{R_{\lambda,q}(r)} \\
     p^{\theta} = \dot{\theta} = \rho^{-2} \sqrt{\Theta_{\lambda,q}(\theta)}\\
     p^{\phi} = \dot{\phi} = \rho^{-2} \Delta^{-1}
     \left[ 2ar + \lambda \left( \rho^{2} - 2r \right)  
\mathrm{cosec}^{2} \theta \right] \\
     p^{t} = \dot{t} = \rho^{-2} \Delta^{-1} \left( \Sigma^{2} -
     2ar \lambda \right)
   \end{split} \end{gather} where as usual $a$ is the angular momentum of
the source (normalised to mass so that $a = 1$ corresponds to the
'extremal' case) and dotted quantities denote differentiation with respect
to some affine parameter. We introduce the metric functions $\Delta =
r^2-2r+a^2$, $\rho^{2} = r^{2} + a^{2} \cos^{2} \theta$, $\Sigma^{2} =
\left( r^{2} + a^{2} \right)^{2} - a^{2} \Delta \sin^{2} \theta$ and the
effective potentials (\citealt{V93}): \begin{gather}
   \label{eqn:2.1.3}
   \begin{split}
       R_{\lambda,q} \left( r \right) = \left[ \left( r^{2} + a^{2}
\right) - a \lambda \right] ^{2}
       - \Delta \left[ q - \left( \lambda - a \right) ^{2} \right] \\
       \Theta_{\lambda,q} \left( \theta \right) = q + a^{2} \cos^{2}
\theta -\lambda^{2} \mathrm{cot}^{2} \theta
     \end{split} \end{gather} Here $\lambda, q$ are constants of motion
which are related to the photons covariant angular and linear momenta (see
\citealt{C83,FP99}). The Kerr metric is both stationary and axisymmetric
so the derived set of geodesic equations (\ref{eqn:2.1.2}) are independent
of both of the coordinates $t$ and $\phi$. The properties of a given
geodesic path (specified by a $\lambda, q$ pair) are completely determined
by the first two of these equations, which can be merged together and
integrated to form a governing equation: \begin{equation}
   \label{eqn:2.1.4}
   \begin{split}
     \pm \int^{r} { \frac{dr}{\sqrt{R_{\lambda,q}(r)}}} =
     \pm \int^{\theta} {\frac{d\theta}{\sqrt
{\Theta_{\lambda,q}(\theta)}}} \end{split} \end{equation} Consider the
radial motion of the geodesic, described by the left-hand side of the
above. We fix the ends of the radial trajectories at $r_{e}$ and $r_{o}$,
which leads us to the general form for radial motion: \begin{equation}
  \label{eqn:2.2.2}
  \begin{split}
    I^{s^{1}_{r} , s^{2}_{r}}_{r, \lambda , q} \left( r_{e} , r_{o}
\right)
    = \pm \int^{r_{o}}_{r_{e}} { \frac{dr}{\sqrt{R_{\lambda,q}(r)}}} \\
    = s^{1}_{r} \left[ \int^{r_{t}}_{r_{e}}
{\frac{dr}{\sqrt{R_{\lambda,q}(r)}}} +
    s^{2}_{r} \int^{r_{t}}_{r_{o}} {\frac{dr}{\sqrt{R_{\lambda,q}(r)}}}
\right]
  \end{split} \end{equation} Here, we have denoted the radial turning
point (either perihelion or aphelion) of the path motion by $r_{t}$ and
$s^{1,2}_{r}$ can take the values $-1,+1$ dependent on the nature of the
path. From the discussion of \cite{W72}, there are no bound null geodesics
that terminate above the horizon, implying that at most there can be one
radial turning point along the geodesic path. There are therefore three
types of radial motion that we must consider: \begin{enumerate} \item From
$r_{e}$ to $r_{o}$ with no radial turning point encountered, implying
either $\left( s^{1}_{r} = +1, s^{2}_{r} = -1 \right)$ if $r_{e} < r_{o}$
or $\left( s^{1}_{r} = -1, s^{2}_{r} = -1 \right)$ if $r_{e} > r_{o}$.
\item From $r_{e}$ inwards to perihelion at $r_{t}$, then outwards to
$r_{o}$, implying that $\left( s^{1}_{r} = -1, s^{2}_{r} = +1 \right)$ for
$r_{t} < r_{e}$, $r_{t} < r_{o}$. \item From $r_{e}$ outwards to aphelion
at $r_{t}$, then inwards to $r_{o}$, implying that $\left( s^{1}_{r} = +1,
s^{2}_{r} = +1 \right)$ for $r_{t} > r_{e}$, $r_{t} > r_{o}$.
\end{enumerate}

For the latitudinal motion we can, in general, have an arbitrary number of
turning points occur along the path (unlike in the radial case). This
requires a more complex expression to describe the contribution of the
latitudinal motion. We introduce the variable $m = \cos \theta$:
\begin{equation}
  \label{eqn:2.2.3}
  \begin{split}
    I^{s^{1}_{m} , s^{2}_{m} , n_{m}}_{m, \lambda , q} \left( m_{e} ,
m_{o} \right)
    =\pm \int^{m_{o}}_{m_{e}} {\frac{dm}{\sqrt {M_{\lambda,q}(m)}}} \\
    = s^{1}_{m} \left[ \int^{m_{+}}_{m_{e}} {\frac{dm}{\sqrt
{M_{\lambda,q}(m)}}}
    + s^{2}_{m} \int^{m_{+}}_{m_{o}} {\frac{dm}{\sqrt {M_{\lambda,q}(m)}}}
\right] \\
    + \left[ 2 n_{m} - s^{1}_{m} s^{2}_{m} \left( 1 + s^{2}_{m} \right)
\right]
    \int^{m_{+}}_{m_{-}} {\frac{dm}{\sqrt {M_{\lambda,q}(m)}}}
  \end{split} \end{equation} The effective potential now takes the form
$M_{\lambda,q} (m) = q + m^{2} \left( a^{2} - \lambda^{2} - q \right) -
a^{2} m^{4}$. In the above, $m_{+,-}$ denote the locations of the
latitudinal turning points determined by solution of $M_{\lambda , q} (m)
= 0$ \citep{RB94}, whilst $n_{m}$ denotes the number of such turning
points through which the path passes and $s^{1,2}_{m}$ can take the values
$-1,+1$. Specifically, the case of $s^{1}_{m}$ is described by two
possibilities: \begin{enumerate} \item If $m_{+} = - m_{-}$, then
$s^{1}_{m} = +1, -1$ dependent on whether the path is initially directed
towards the 'north' or 'south' poles of the hole respectively. \item If
$m_{+} \ne - m_{-}$, then $s^{1}_{m}$ is in the direction of the closest
of $m_{+,-}$ to $m_{e}$ \end{enumerate} The sign of $s^{2}_{m}$ is
determined by the number of turning points, $n_{m}$ through which the
geodesic passes, $s^{2}_{m} = -1^{n_{m}}$.

Geodesic paths are therefore described by: \begin{equation}
  \label{eqn:2.2.4}
    I^{s^{1}_{r} , s^{2}_{r}}_{r, \lambda , q} \left( r_{e} , r_{o}
\right) -
    I^{s^{1}_{m} , s^{2}_{m} , n_{m}}_{m, \lambda , q} \left( m_{e} ,
m_{o} \right) = 0 \end{equation} This can be solved analytically in terms
of elliptic functions, which is much more efficient than numerical
solutions. \cite{RB94} tabulate these functions to determine the observed
coordinate latitude, $\theta_{obs}$ for photons with momenta $\left(
\lambda, q \right)$ emitted from a given radius and latitude which arrive
at the observers' radial coordinate. Alternatively, \cite{APHD,C98} fix
one end of the radial path at infinity with some inclination and the
coordinates at which the other end crosses the equatorial plane. An
additional method, due to \cite{V93} fixes the ends of the photon paths at
the required coordinates, and a minimisation technique is applied to
determine valid $\left( \lambda, q \right)$ pairs for a given number of
orbits of the black hole.

Our approach combines aspects of those described by \cite{RB94} and
\cite{V93}. We invert the reformulated governing equation to obtain the
observed co-ordinate latitude of the geodesic, using the technique
described by \cite{RB94}: \begin{equation}
  \label{eqn:2.2.5}
  \begin{split}
    \int^{m_{+}}_{m_{o}} {\frac{dm}{\sqrt {M_{\lambda,q}(m)}}}
    = s^{1}_{m} s^{2}_{m} I^{s^{1}_{r} , s^{2}_{r}}_{r, \lambda , q}
\left( r_{e} , r_{o} \right) \\
    - s^{2}_{m} \int^{m_{+}}_{m_{e}} {\frac{dm}{\sqrt {M_{\lambda,q}(m)}}}
\\
    - s^{1}_{m} s^{2}_{m} \left[ 2 n_{m} - s^{1}_{m} s^{2}_{m} \left( 1 +
s^{2}_{m} \right) \right]
    \int^{m_{+}}_{m_{-}} {\frac{dm}{\sqrt {M_{\lambda,q}(m)}}}
  \end{split} \end{equation} We apply the properties of the effective
potentials described by \cite{V93} to dramatically reduce 
the scale of the calculation by analytically restricting the
range of $\lambda$ and $q$ to those which can escape to infinity. Then
we search this range for those paths which contribute to a given image
order at the required observed inclination. 
These geodesics are then projected to form
an image of the system on the observers sky, which is then used to
determined the measured flux (described in \citealt{BD04}). The code
therefore allows the fast calculation of geodesics linking any two points
in the spacetime that make a specified number orbits of the black hole.

\subsection{The Zeroes of the Effective Potentials}

\begin{figure*}
  \leavevmode
  \begin{center}
  \begin{tabular}{ccc}
  \includegraphics[width=0.3\textwidth]{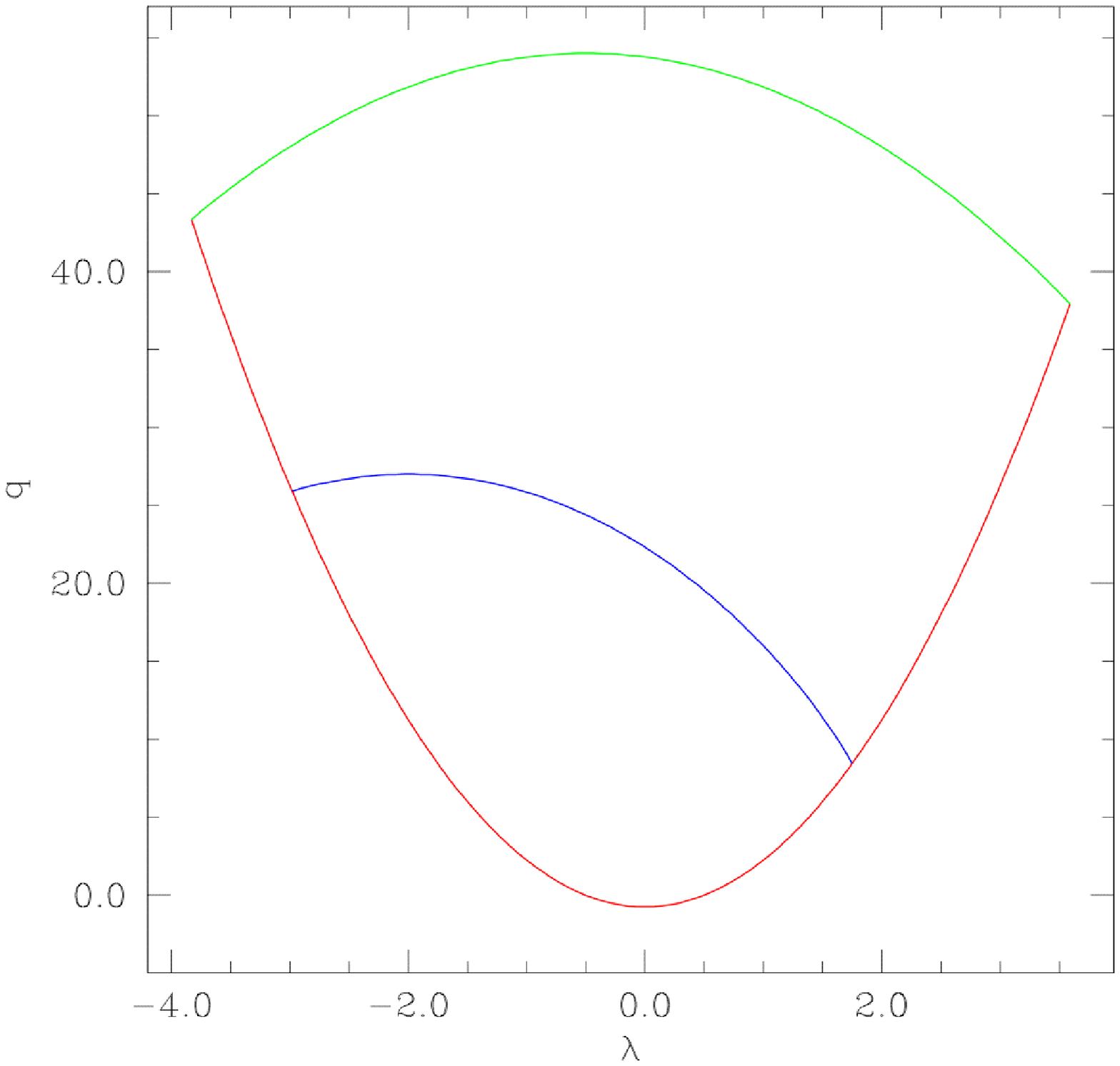}
  &
  \includegraphics[width=0.3\textwidth]{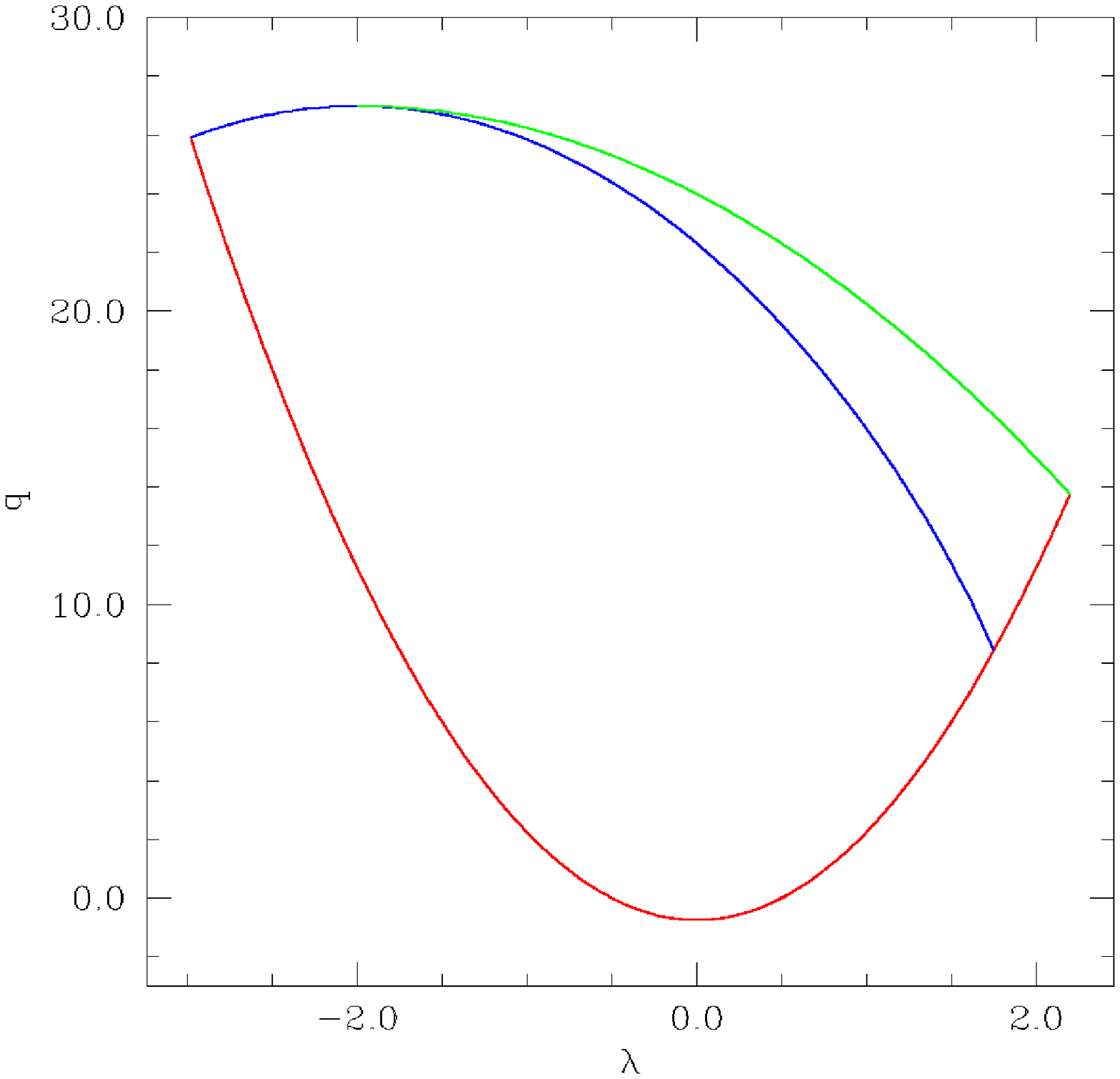}
  &
   \includegraphics[width=0.3\textwidth]{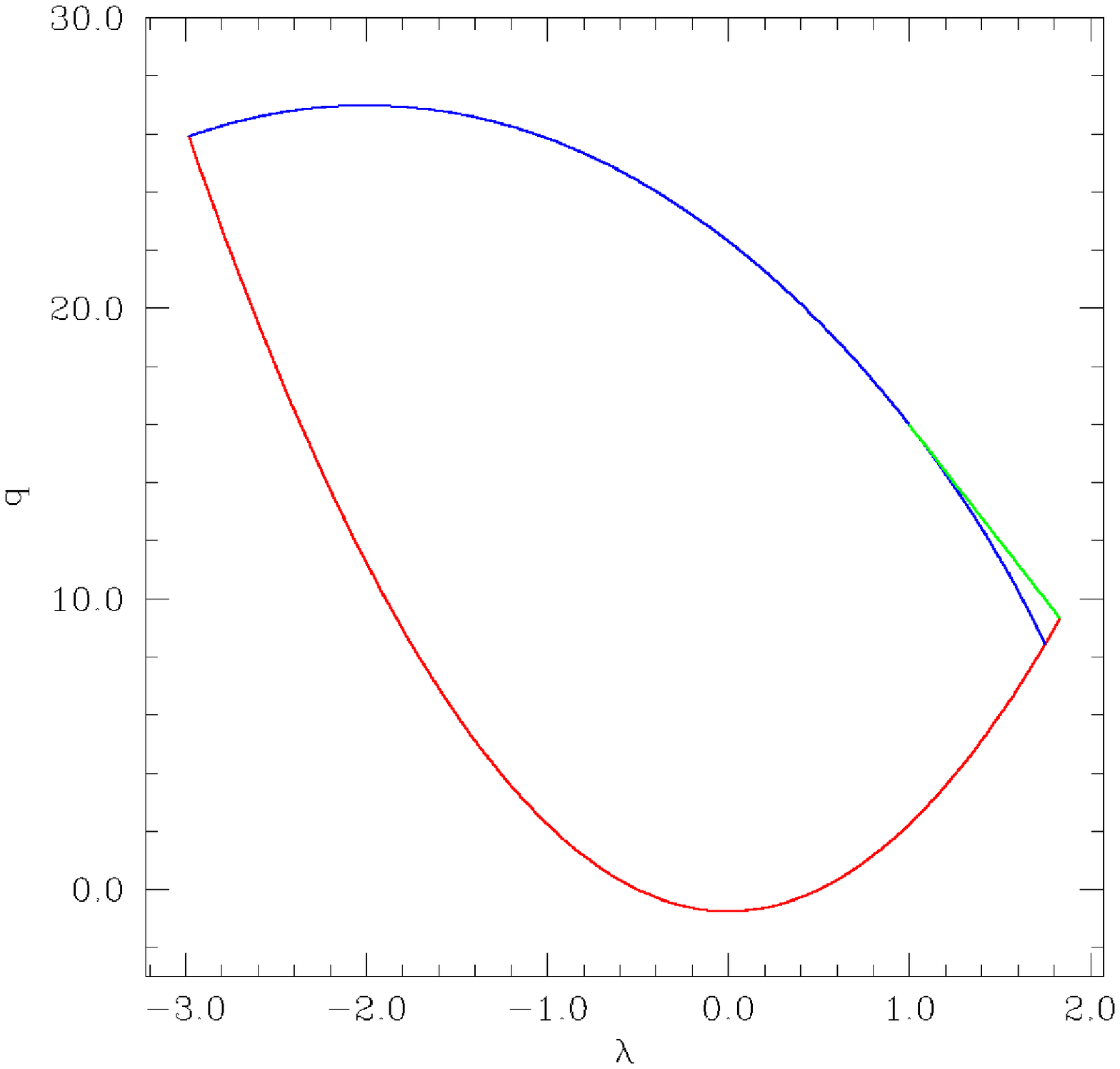}
  \end{tabular}
  \caption{Boundary curves on the $(\lambda, q)$-plane for an extreme
($a=1$) Kerr black hole with $r_{o} = \infty$, $\theta_{e} = \pi / 2$ and
$\theta_{o} = 30^{\circ}$. The graph of $q_{m,a} \left( \lambda \right)$
is shown by the red curve, $q_{r,a} \left( \lambda \right)$ by the green
and $q_{c} \left( r_{c} \right), \lambda_{c} \left( r_{c} \right)$ by the
blue. On the left-hand panel, the emitter is located at $r_{e} = 6 r_{g}$,
such that $r_{e} > r^{(-)}_{ph}$ and hence the valid region is bounded by
$q_{r,a} \left( \lambda \right)$ from above and $q_{m,a} \left( \lambda
\right)$ from below, whilst the $q_{c} \left( r_{c} \right), \lambda_{c}
\left( r_{c} \right)$ describes the line below which a photon directed
radially inward is inevitably captured by the hole. In the central panel,
the emitter is now located at $r_{e} = 3 r_{g}$, such that $r^{(-)}_{ph}
\le r_{e} \le r^{(-)}_{ph}$. The parameter space is again bounded by
$q_{m,a} \left( \lambda \right)$ from below, however in this case, the
upper limit is given by $q_{c} \left( r_{c} \right), \lambda_{c} \left(
r_{c} \right)$ for $\lambda \le \lambda_{c} \left( r_{e} \right)$ and by
$q_{r,a} \left( \lambda \right)$ otherwise. Finally, we see that for
$r_{e} = 2 r_{g}$ (right-hand panel), these boundaries still apply, even
though the graph of $q_{r,a} \left( \lambda \right)$ is now concave (true
for all $r \le 2 r_{g}$).}
  \label{fig:2.2.1}
  \end{center} \end{figure*}

To obtain solutions of the reformulated governing equation
(\ref{eqn:2.2.4}), we turn to the tables of elliptic integrals provided by
\cite{RB94}, as modified by \cite{APHD}. These tables, when combined with
appropriate numerical techniques allow us, in principle, to determine the
geodesics that link an arbitrary emitter, $\left( r_{e}, m_{e} \right)$
and observer, $\left( r_{o}, m_{o} \right)$.

In practice, however, this calculation is far from trivial. By specifying
the locations of the emitter and receiver, we have placed definite
restrictions on the values of the angular momentum parameters, $( \lambda,
q)$ for which geodesic motion between these two locations is even
possible. The geodesic motion is dependent on the square root of the two
effective potentials, $R_{\lambda,q} (r)$, $M_{\lambda,q} (m)$, which
requires that these functions remain positive definite at all points along
the path. If, at any point on the geodesic, this requirement is broken,
then a potential barrier is necessarily formed and no such motion is
possible. \cite{V93} has shown that these requirements can be expressed in
terms of the interplay of a set of algebraic functions, which we consider
further here. Note that the application of these functions enables us to
provide tight limits on the region of parameter space which must be
considered in the calculation and hence hugely reduce the scale of the
calculation.

We begin by introducing: \begin{gather*}
  \underline{r} = \min \left( r_{e} , r_{o} \right) ; \;\;
  \overline{r} = \max \left( r_{e} , r_{o} \right) \\
  \underline{m} = \min \left( \left| m_{e} \right| , \left| m_{o} \right|
\right) ; \;\;
  \overline{m} = \max \left( \left| m_{e} \right| , \left| m_{o} \right|
\right) \end{gather*} The condition that no potential barrier exists
between the emitter and observer can be re-expressed mathematically as:
\begin{equation}
  \label{eqn:2.3.1}
    R_{\lambda,q} \left( \underline{r} \le r \le \overline{r} \right) \ge
0 ; \;\;
    M_{\lambda,q} \left( \underline{m} \le \left| m \right| \le
\overline{m} \right) \ge 0 \end{equation} Since the effective potentials
are linear in $q$, we can express these requirements as: \begin{equation}
  \label{eqn:2.3.2}
  q \le q_{r,a} \left( \lambda \right) ; \;\;
  q \ge q_{m,a} \left( \lambda \right) \end{equation} Here:
\begin{equation}
  \label{eqn:2.3.3}
  \begin{split}
    q_{r,a} \left( \lambda \right) = \frac{\left[ r^{2} + a \left( a -
\lambda \right) \right] ^{2}}{\Delta}
    + \left( \lambda -a \right)^{2} \\
    q_{m,a} \left( \lambda \right) = \frac{m^{2}}{1 - m^{2}} \lambda^{2} -
a^{2} m^{2}
  \end{split} \end{equation} Physically, these curves correspond to the
locus of points on the $\left( \lambda, q \right)$ plane for which the
given co-ordinate, $\left( r , m \right)$ is a zero of the associated
effective potential. For the latitudinal motion, it is found that there
are two cases that we must consider (taking $\underline{m}$ as the
pericentre of the latidunal motion and similarly $\overline{m}$ is the
apocentre): \begin{equation}
  \label{eqn:2.3.4}
  q \ge \begin{cases}
  q_{\underline{m},a} \left( \lambda \right)&
  \left| \lambda \right| < a \sqrt{ 1 - \underline{m}^{2}} \sqrt{ 1 -
\overline{m}^{2}} \\
  q_{\overline{m},a} \left( \lambda \right)&
  \left| \lambda \right| \ge a \sqrt{ 1 - \underline{m}^{2}} \sqrt{ 1 -
\overline{m}^{2}}
  \end{cases} \end{equation}

The description of the radial motion is more complex, due to the existence
of the unstable photon orbits \cite{C83}. These orbits are described by
the existence of a further set of zeroes of the radial effective
potential, $R_{\lambda,q} (r) = 0$ that is subject to the additional
constraint $\partial_{r} R_{\lambda,q} \left( r \right) = 0$. These
conditions yield a pair of parametric equations describing a critical
curve on the $\left( \lambda, q \right)$ plane, which define the apparent
angular size of the black hole: \begin{equation}
  \label{eqn:2.3.5}
  \begin{split}
    \lambda_{c} = \frac{1}{a \left( r_{c} - 1 \right)} \left( r^{2}_{c} -
a^{2} - r_{c} \Delta_{c} \right) \\
    q_{c} = \frac{r^{3}_{c}}{a^{2} \left( r_{c} - 1 \right)^{2}}
    \left[ 4 \Delta_{c} - r_{c} \left( r_{c} - 1 \right)^{2} \right]
  \end{split} \end{equation} The range of values of $r_{c}$ is given by
considering the solutions of $4 a^{2} = r_{c} \left( r_{c} - 3
\right)^{2}$ (corresponding to the radii of unstable photon orbits, either
direct or retrograde, in the equatorial plane). Following \cite{C83}, we
denote these radii by $r^{(+)}_{ph}$ and $r^{(-)}_{ph}$ and so
$r^{(+)}_{ph} \le r_{c} \le r^{(-)}_{ph}$.

The second complication to this discussion stems from \cite{W72}.
Specifically, for any particle, orbits with $E^{2} > \delta^{2}$ (where
$E^{2}$ denotes particle energy; $\delta^{2} = 1, 0$ particle mass), are
unbound, whilst orbits with $E^{2} < \delta^{2}$ are bound. Since photons
are massless, this implies that the only bound orbits are those with
negative $E^{2}$ (and so must terminate behind the horizon), whilst all
others must be unbound (except for a set of unstable, circular orbits). We
cannot therefore (unlike in the previous discussion) consider trajectories
with $\underline{r}$ as pericentre and $\overline{r}$ as apocentre.

\begin{figure*}
  \leavevmode
  \begin{center}
  \begin{tabular}{cccccc}
  \includegraphics[width=0.14\textwidth]{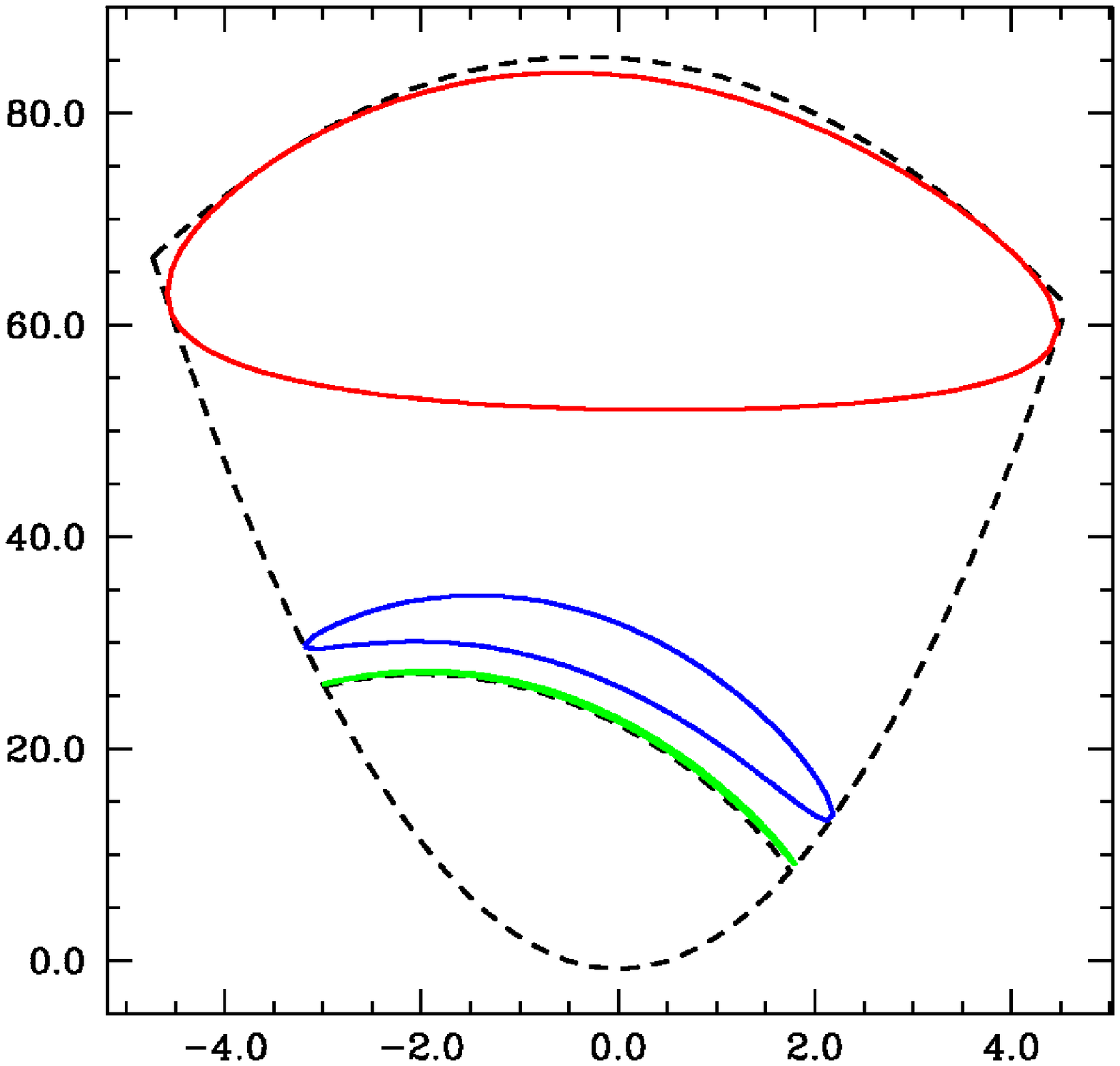}
  &
  \includegraphics[width=0.14\textwidth]{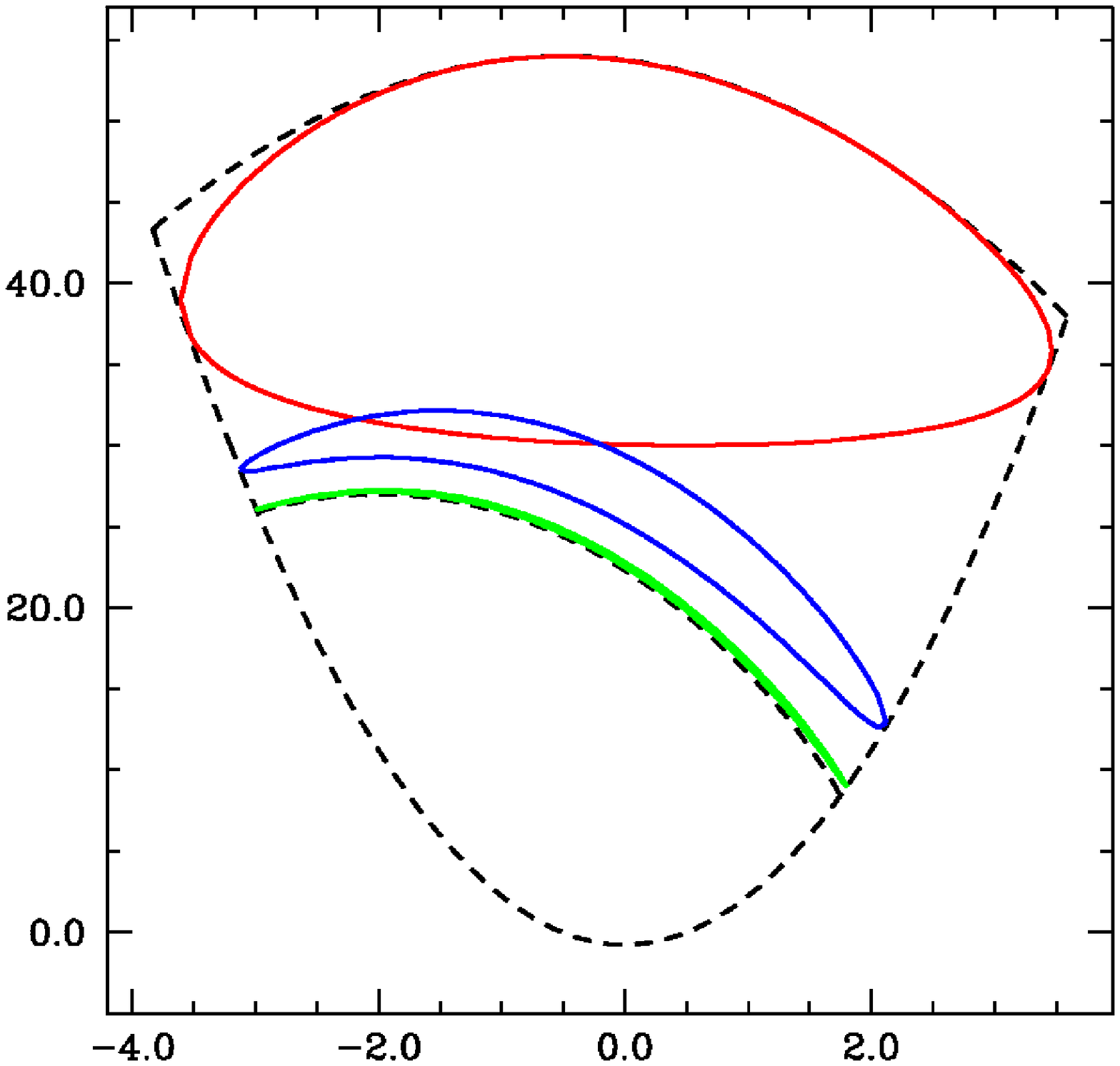}
  &
  \includegraphics[width=0.14\textwidth]{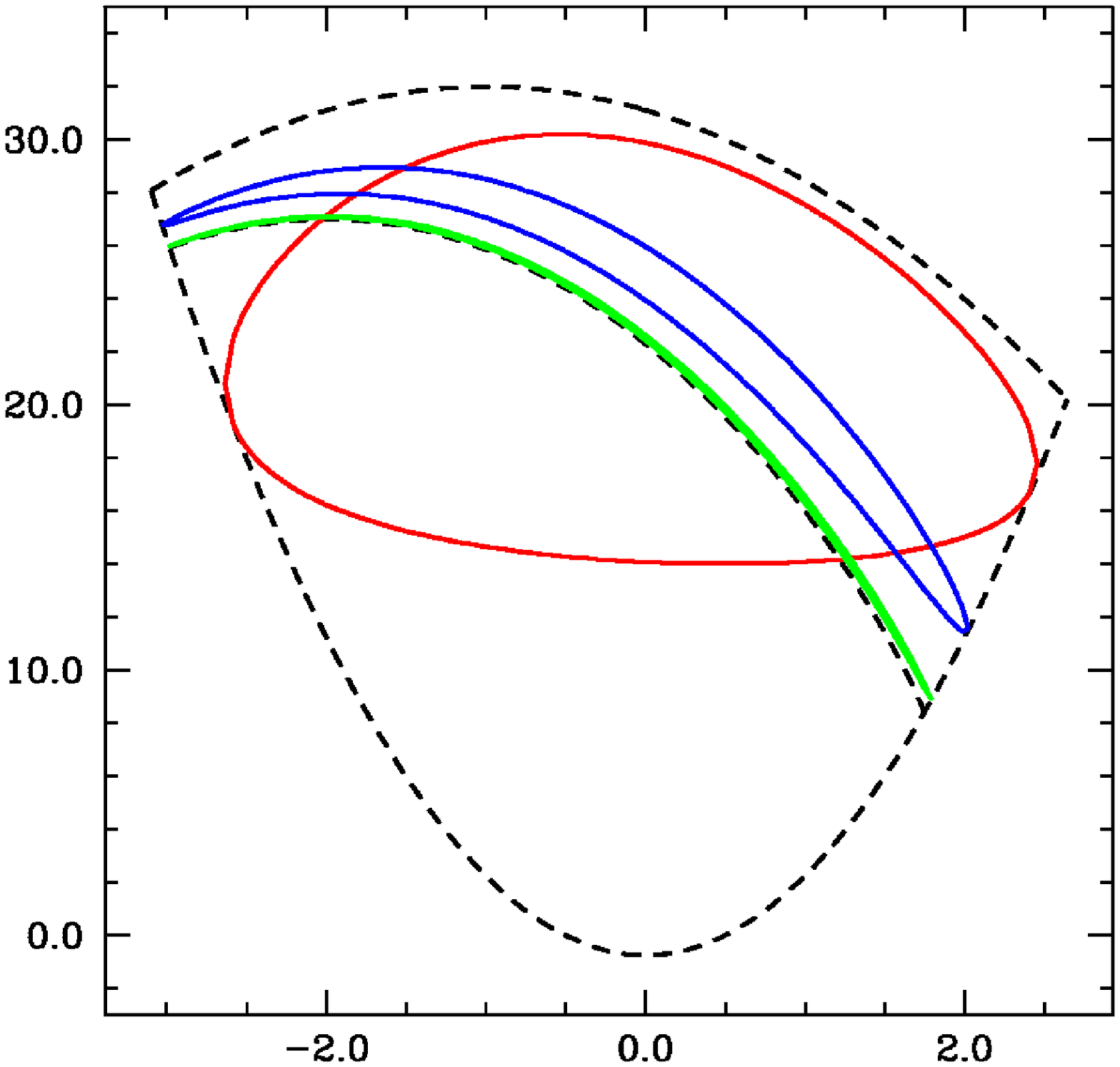}
  &
  \includegraphics[width=0.14\textwidth]{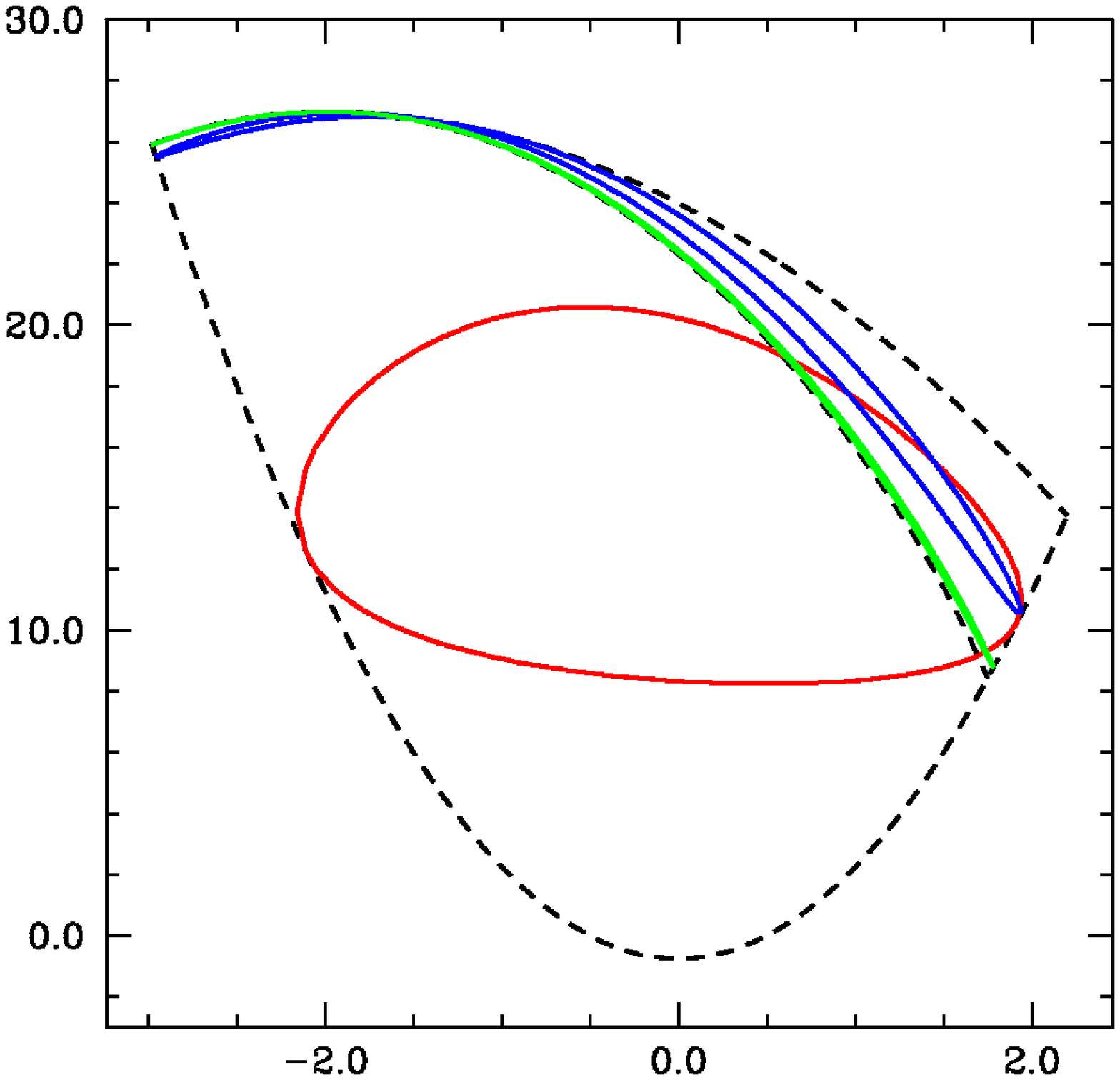}
  &
  \includegraphics[width=0.14\textwidth]{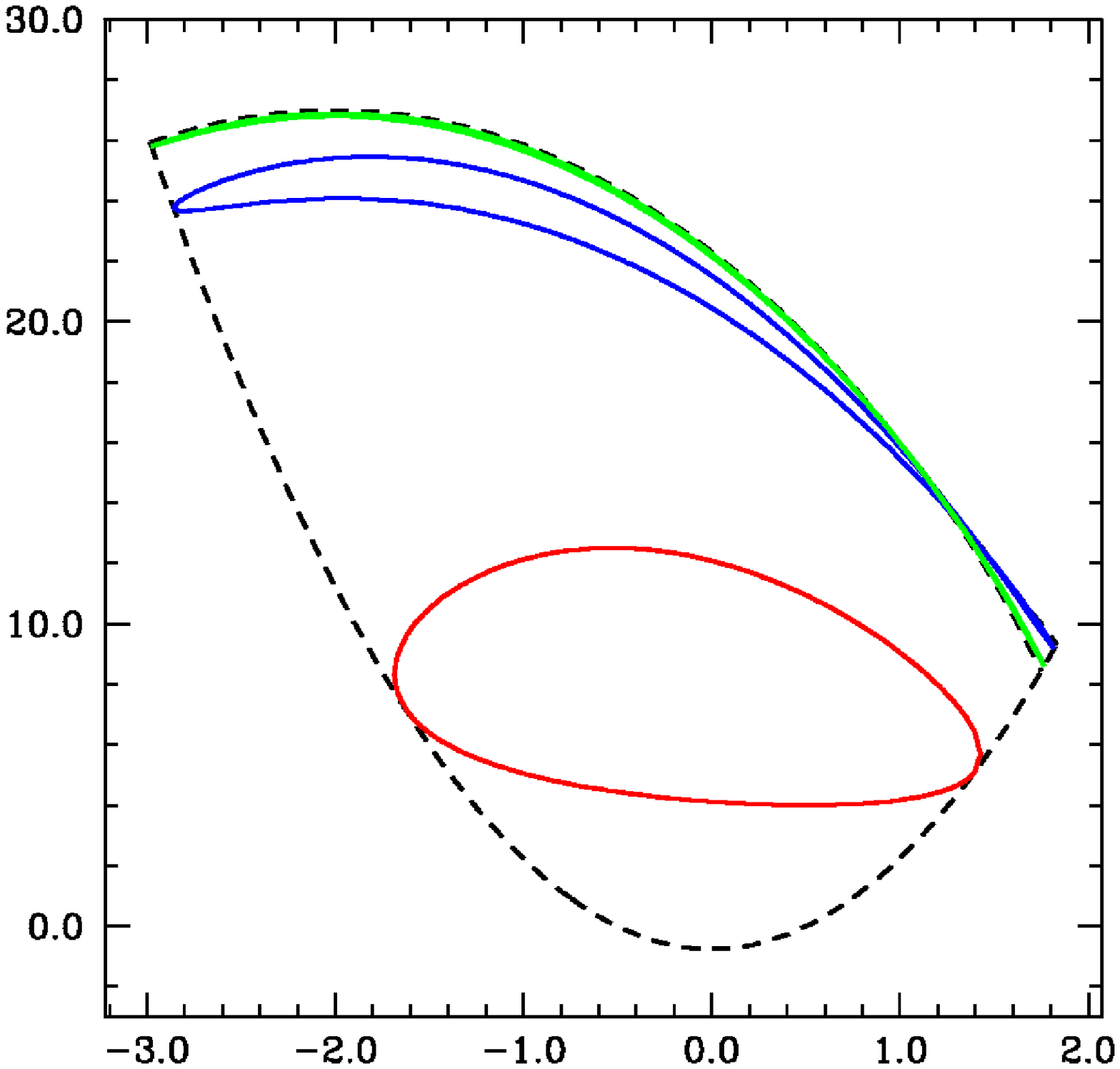}
  &
  \includegraphics[width=0.14\textwidth]{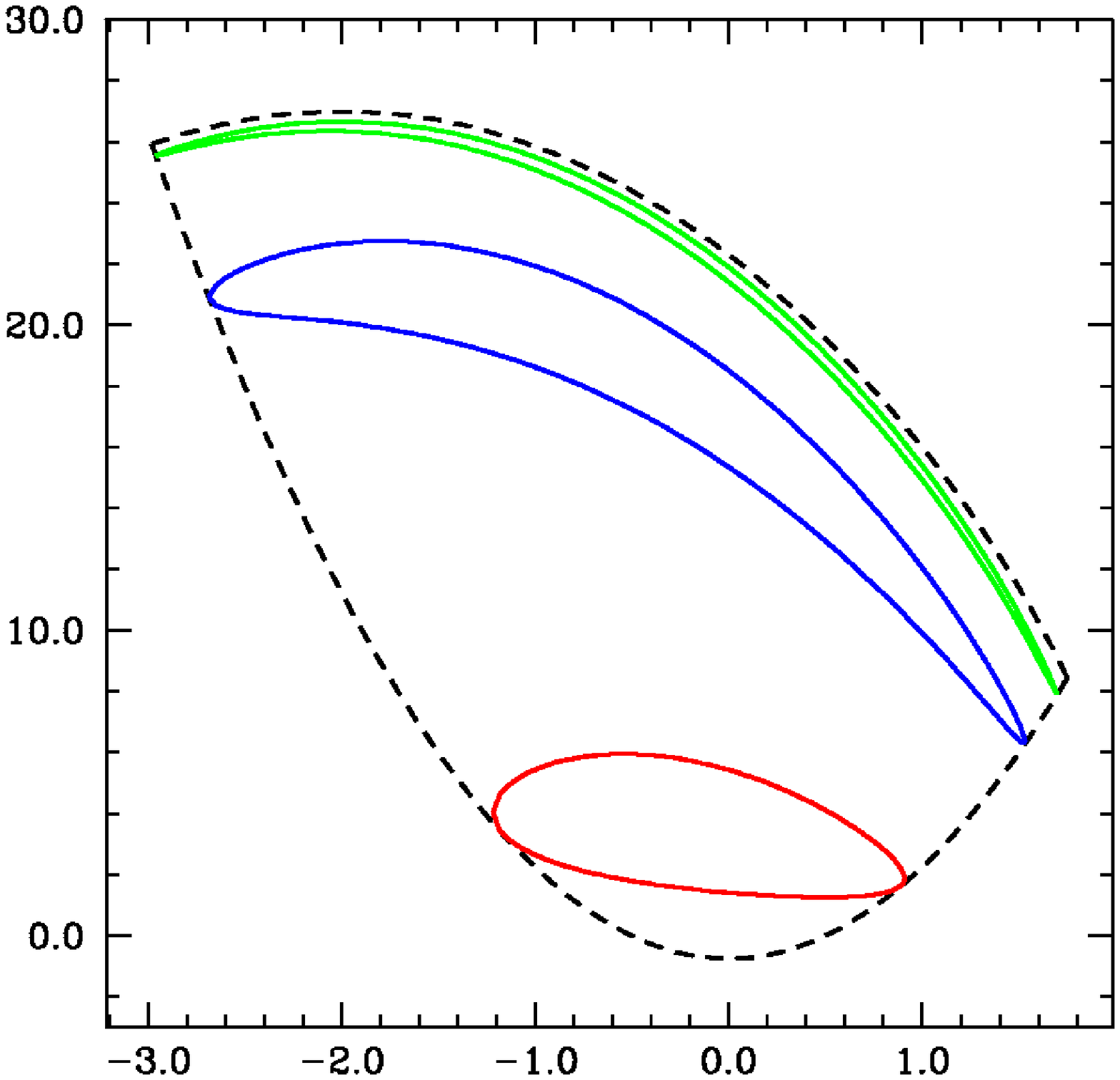}
  \\
  \includegraphics[width=0.14\textwidth]{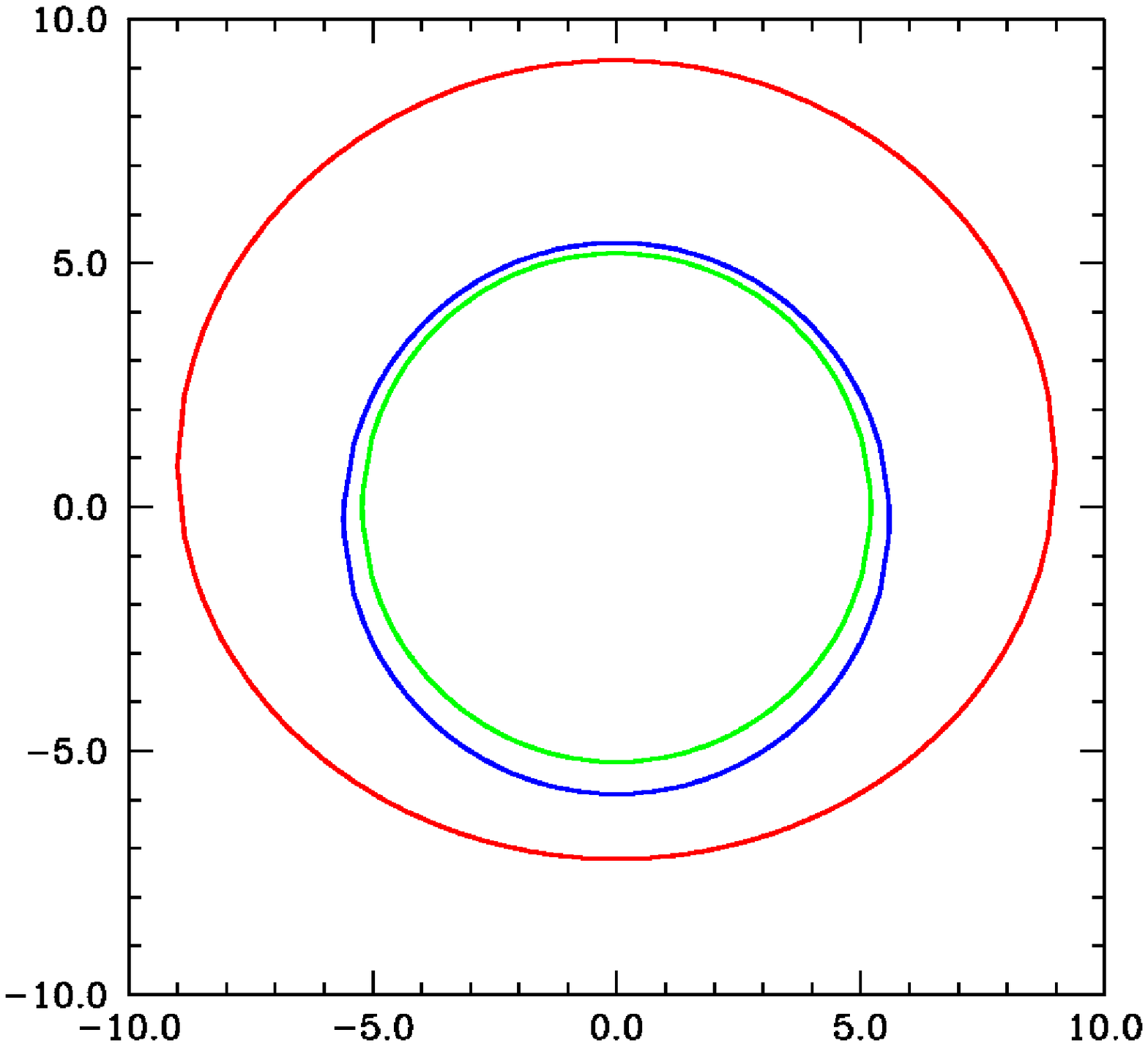}
  &
  \includegraphics[width=0.14\textwidth]{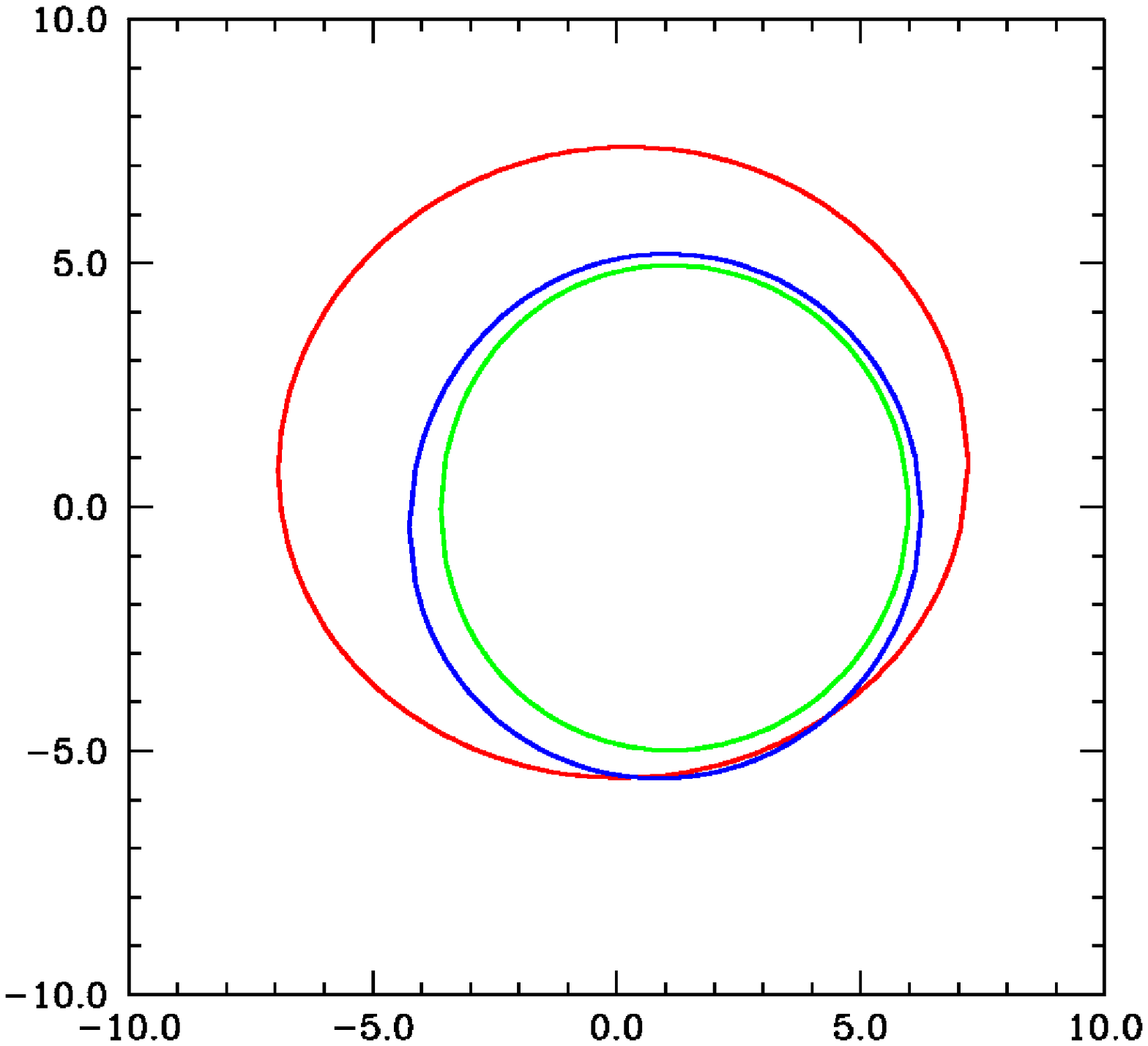}
  &
  \includegraphics[width=0.14\textwidth]{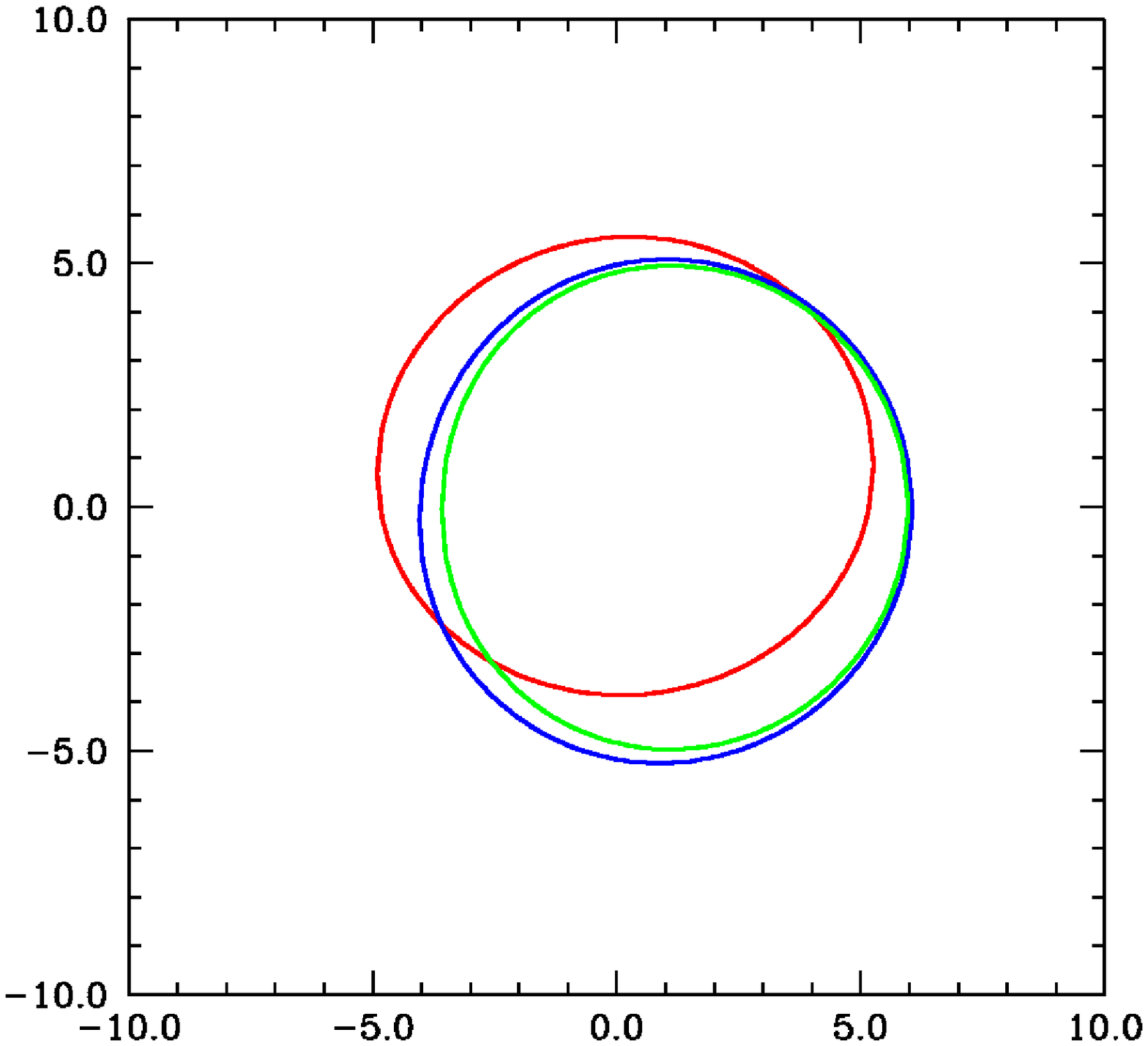}
  &
  \includegraphics[width=0.14\textwidth]{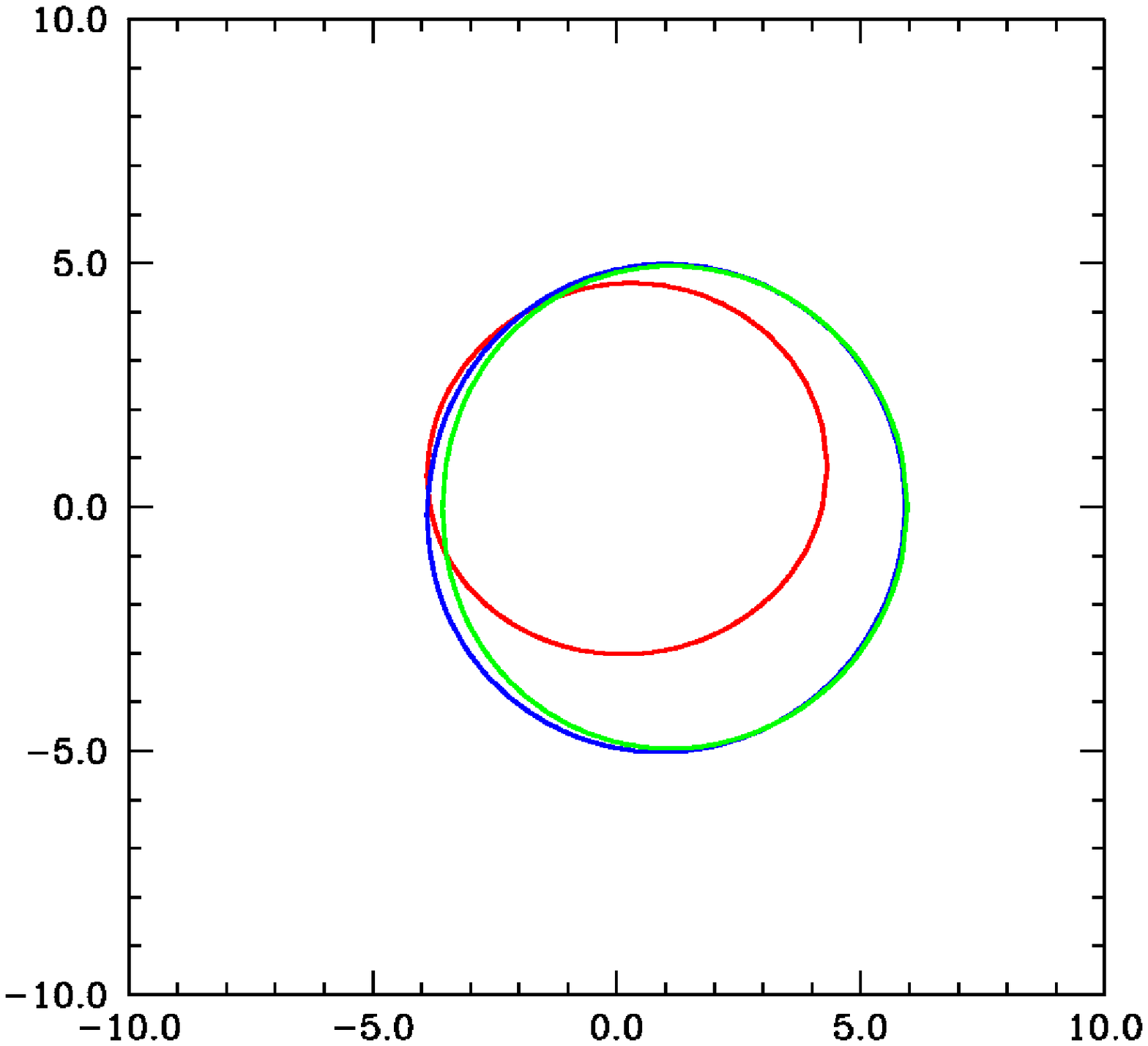}
  &
  \includegraphics[width=0.14\textwidth]{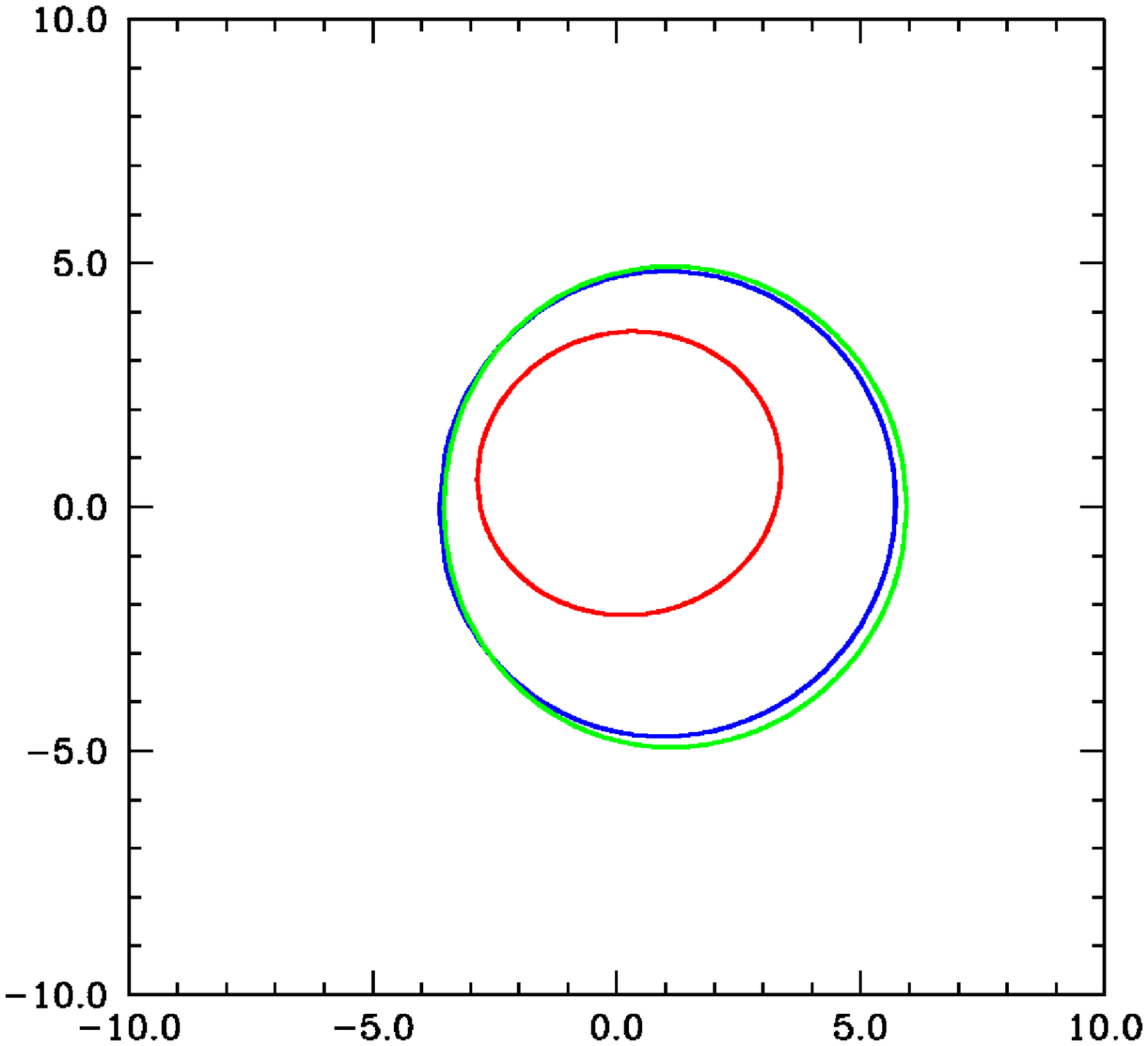}
  &
  \includegraphics[width=0.14\textwidth]{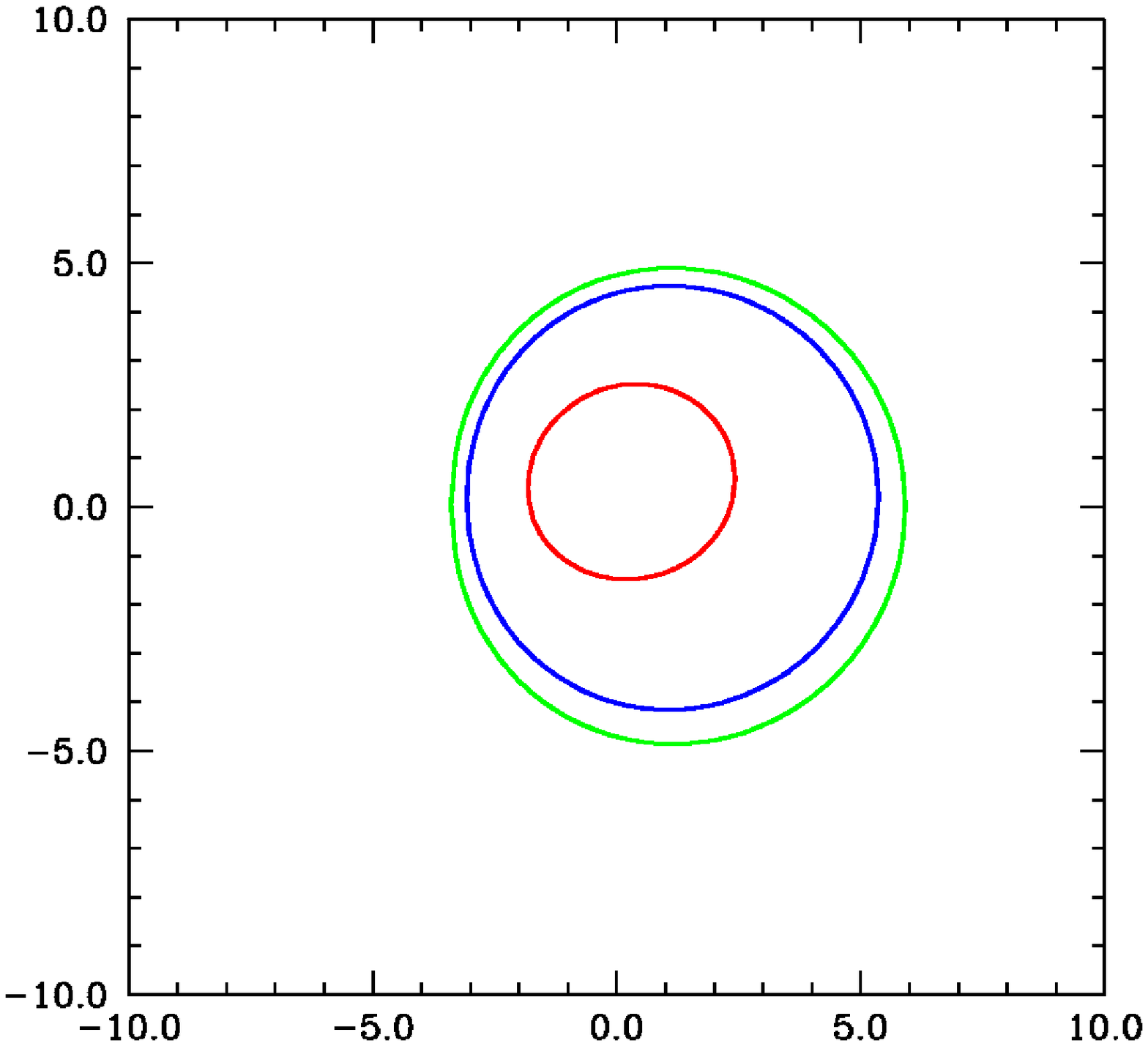}
  \end{tabular}
  \caption{Closed geodesic loops on the $(\lambda, q)$-plane (top row) and
their projection onto the observers sky (the $\alpha, \beta$ plane, bottom
row) for an extreme ($a=1$) Kerr black hole with $r_{o} = \infty$,
$\theta_{e} = \pi / 2$ and $\theta_{o} = 30^{\circ}$. Geodesics with $N =
0$ are shown as red loops, those with $N=1$ as blue loops and $N=2$ as
green loops. Also shown for references on the $(\lambda, q)$-plane
representation are the parameter space boundaries (black dashed lines).
From left to right, the each panel has $r_{e} = 8, 6, 4, 3, 2, 1 r_{g}$.
Note the change in the relative locations of the loops as the emitter
moves inwards - for $r_{e} = 8r_{g}$, the loops for each image order are
completely detached. However, as the emitter moves in to $6r_{g}$, the
$N=0,1$ loops overlap and by the time the emitter reaches $4r_{g}$, the
$N=0$ loop overlaps both the $N=1$ and $N=2$ loop. At $3 r_{g}$, the $N=0$
loop again crosses the $N=1,2$ loops, however in this case the $N=1$ and
$N=2$ loops now touch. Moving the emitter inward to $2 r_{g}$, the $N=0$
loop detaches itself and moves inside the $N=1,2$ loops, which cross.
Moving the emitter still further inward, to just above $r_{+}$, we now see
that $N=1,2$ loops have detached, with the $N=1$ loop now occurring
between the $N=0,2$ loops.}
  \label{fig:2.3.1}
  \end{center} \end{figure*}

\begin{figure*}
  \leavevmode
  \begin{center}
  \begin{tabular}{cccccc}
  \includegraphics[width=0.14\textwidth]{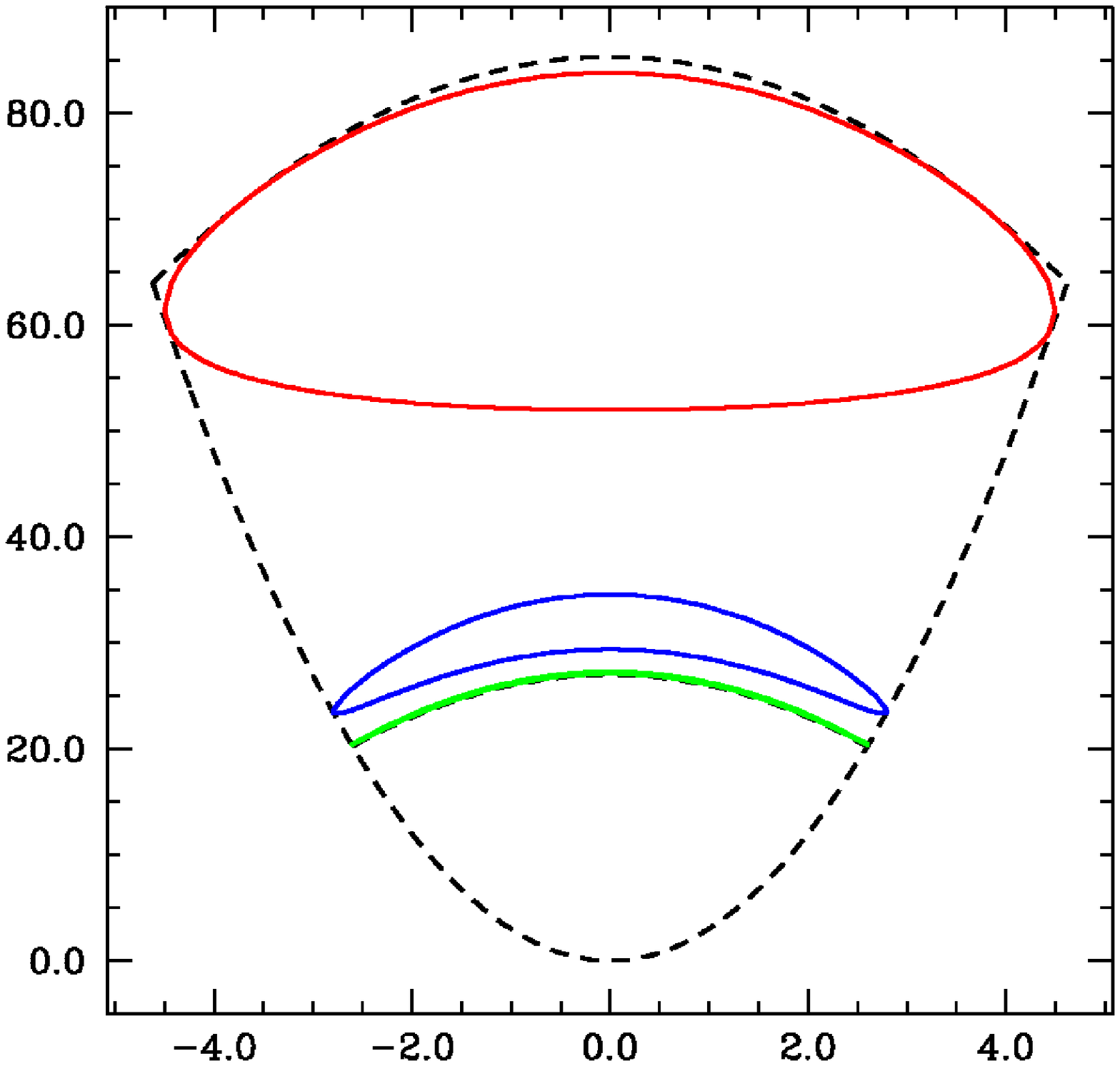}
  &
  \includegraphics[width=0.14\textwidth]{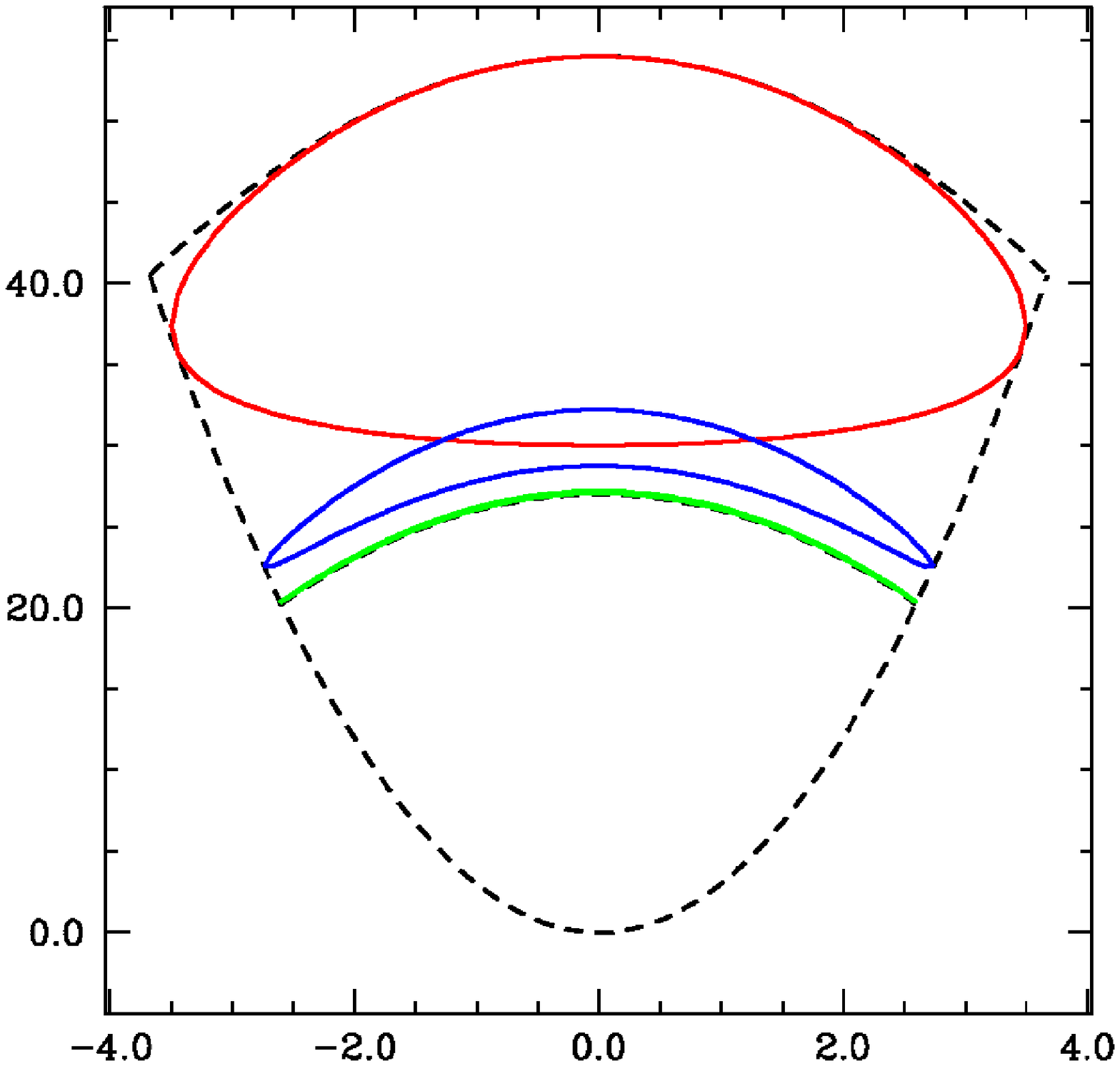}
  &
  \includegraphics[width=0.14\textwidth]{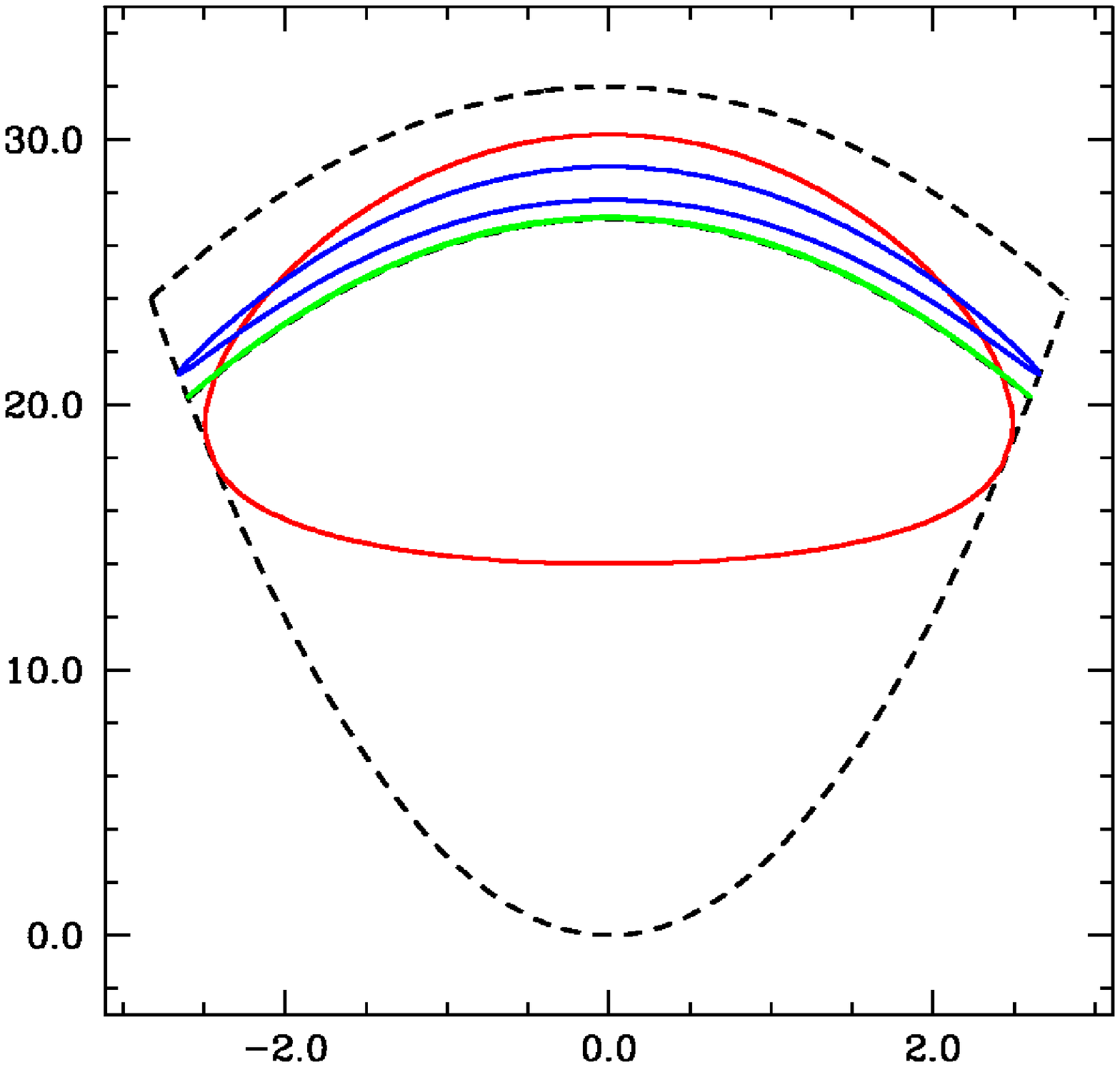}
  &
  \includegraphics[width=0.14\textwidth]{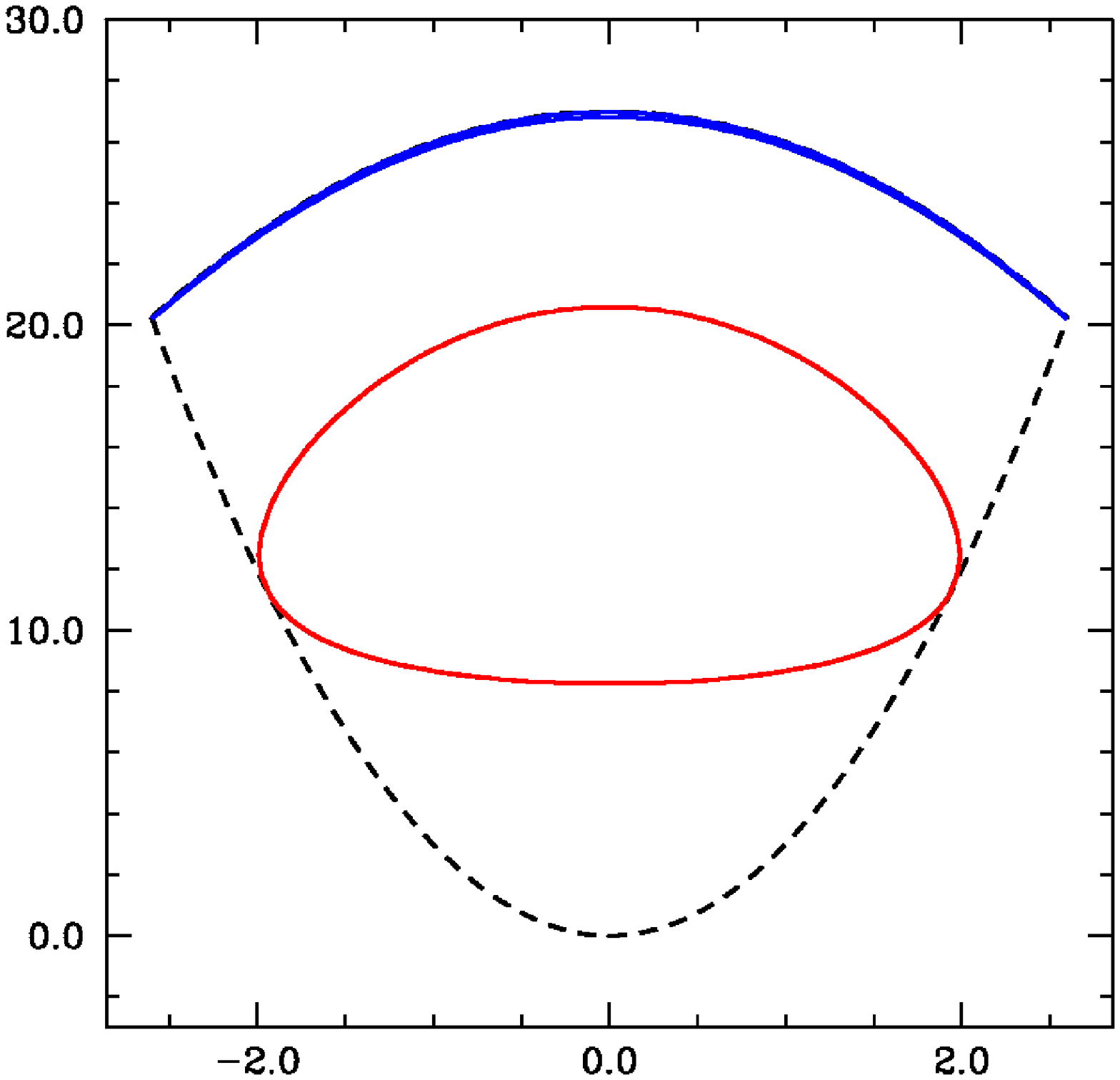}
  &
  \includegraphics[width=0.14\textwidth]{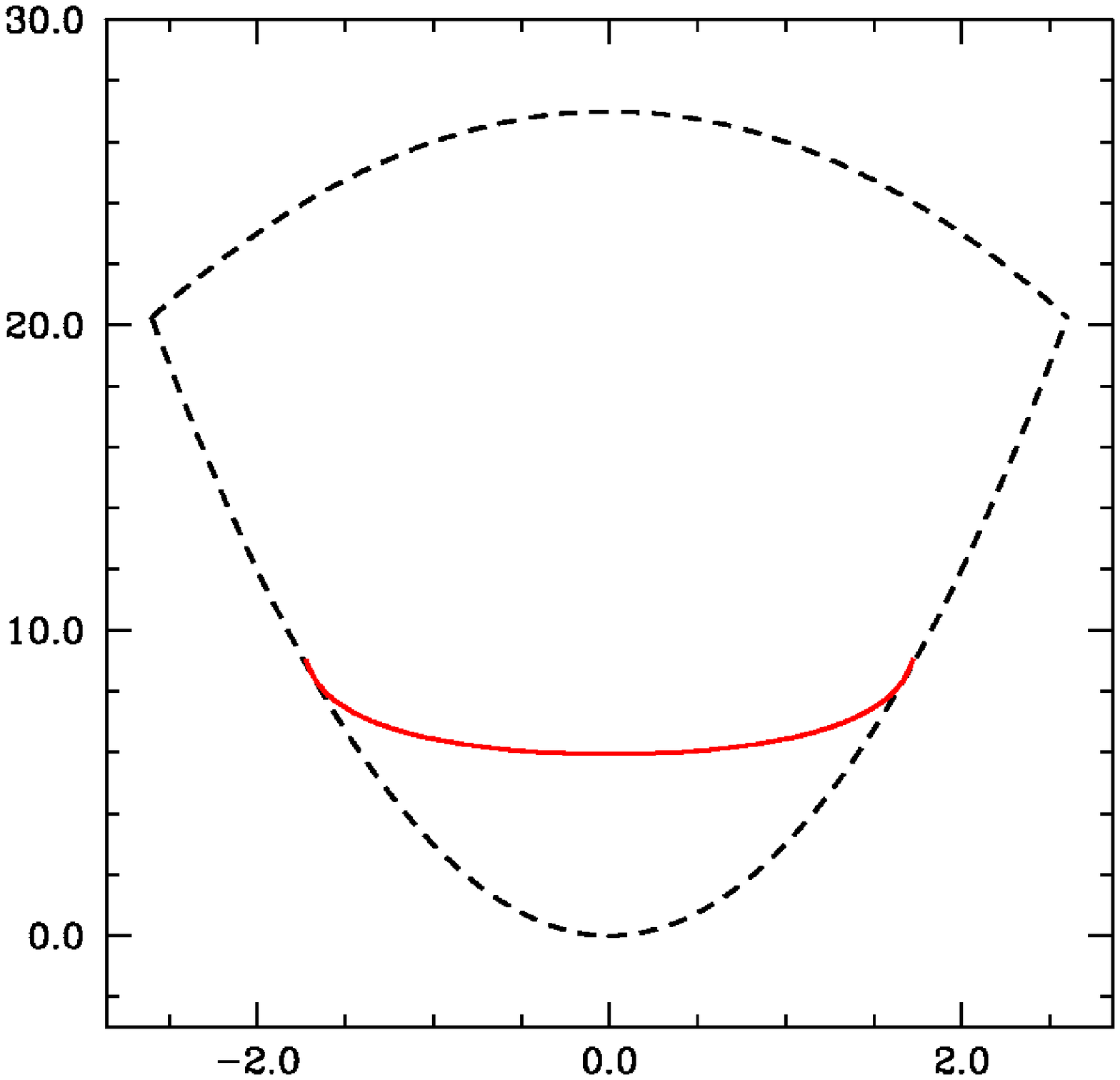}
  &
  \includegraphics[width=0.14\textwidth]{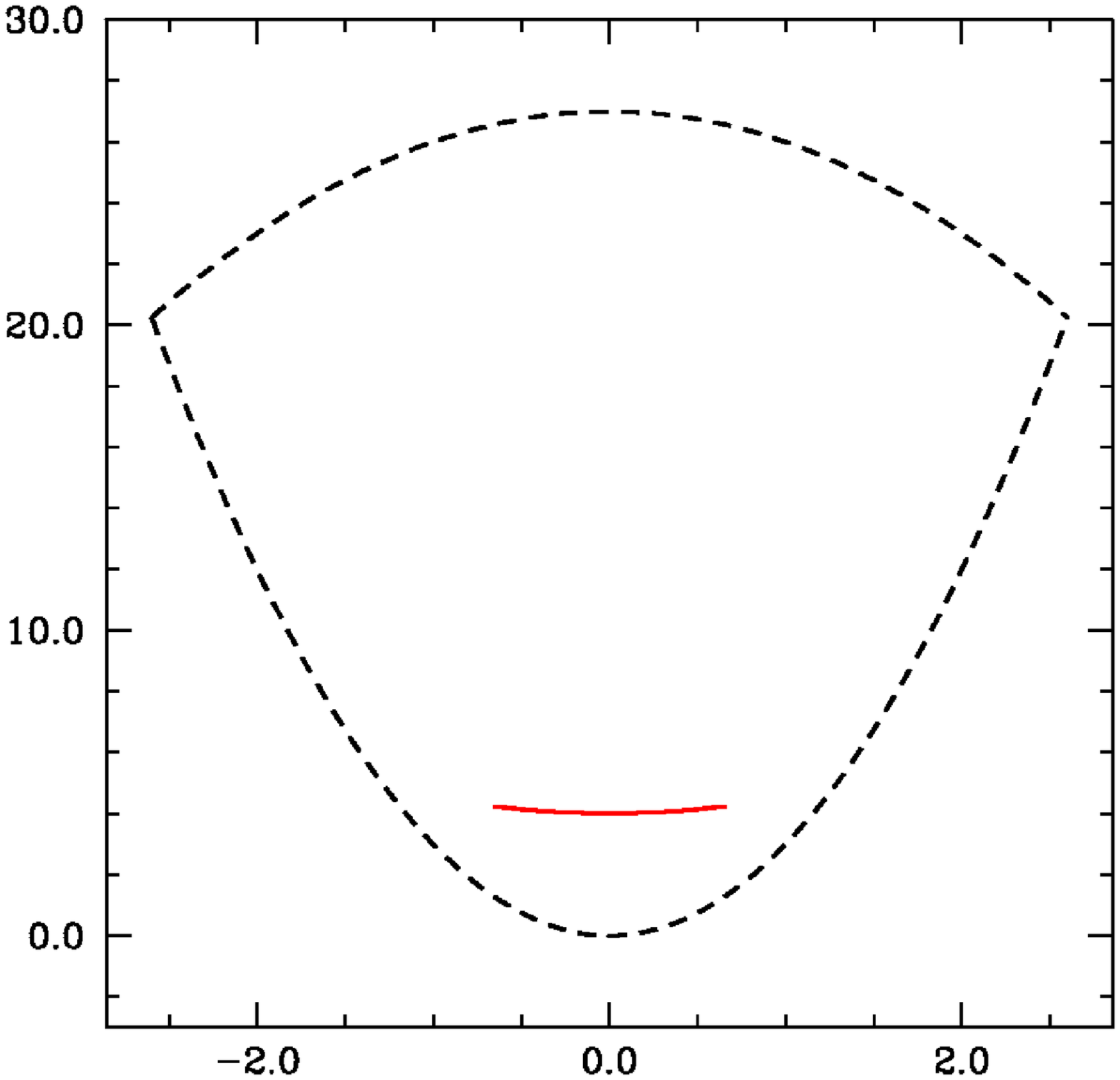}
  \\
  \includegraphics[width=0.14\textwidth]{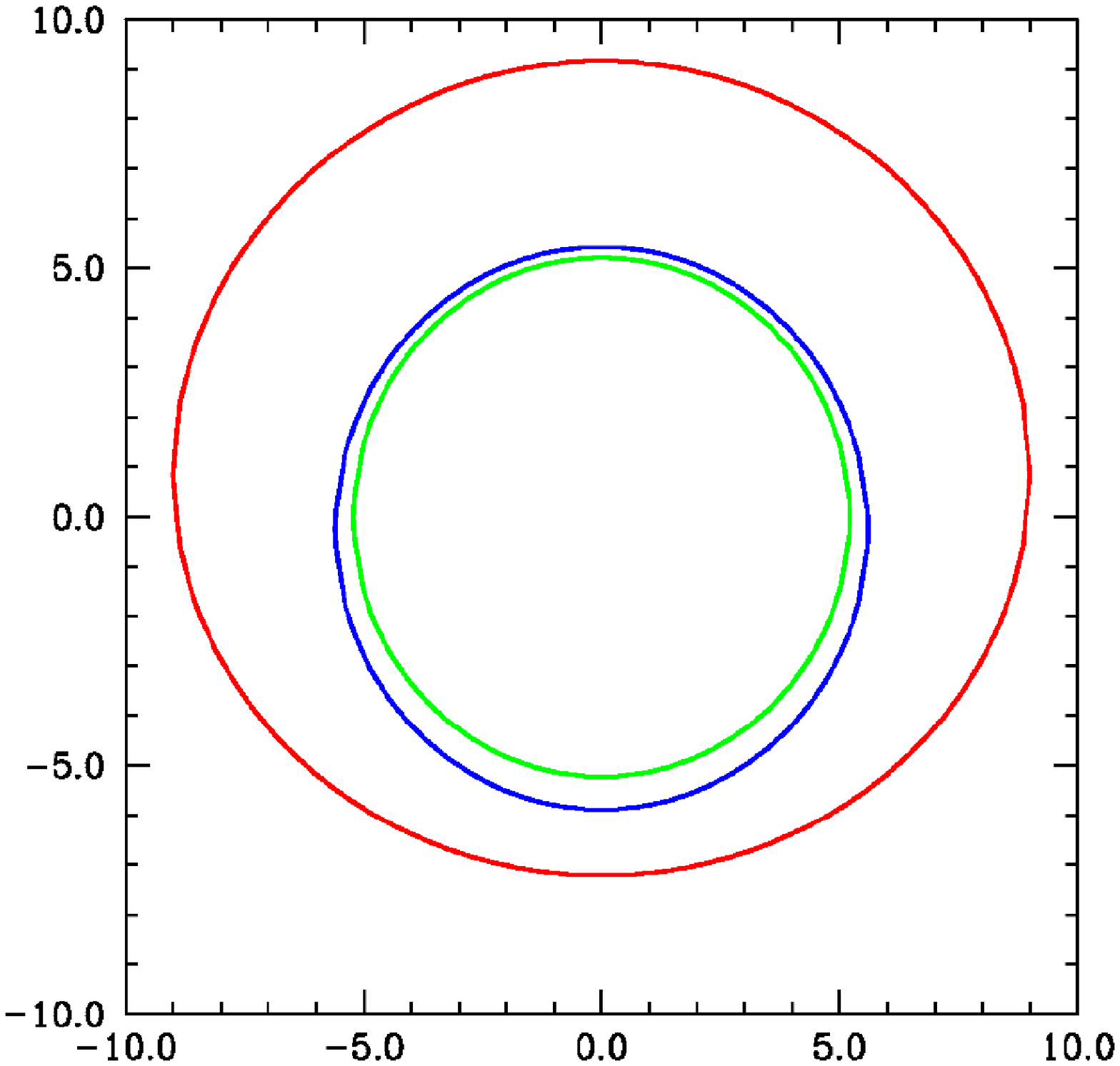}
  &
  \includegraphics[width=0.14\textwidth]{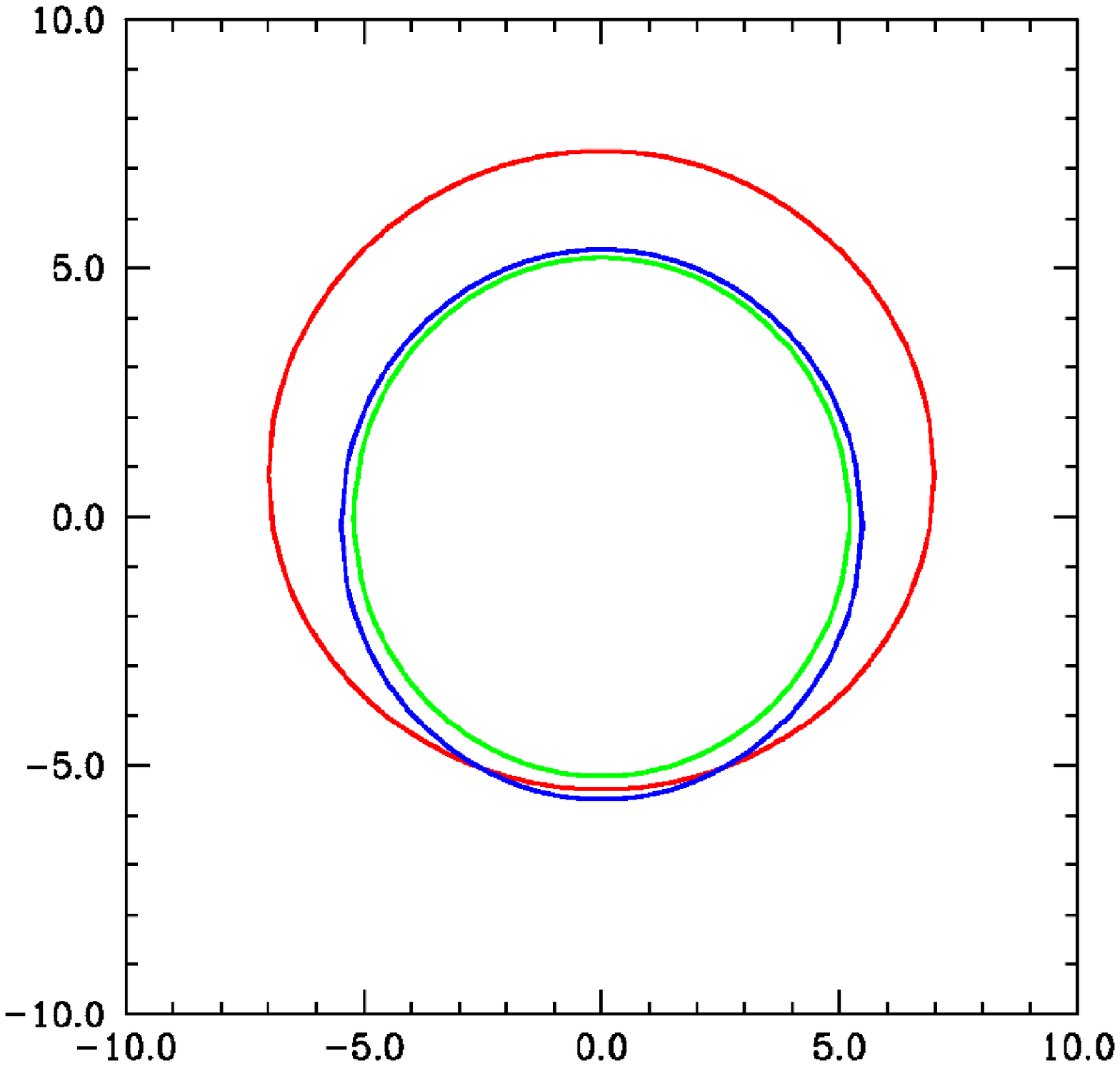}
  &
  \includegraphics[width=0.14\textwidth]{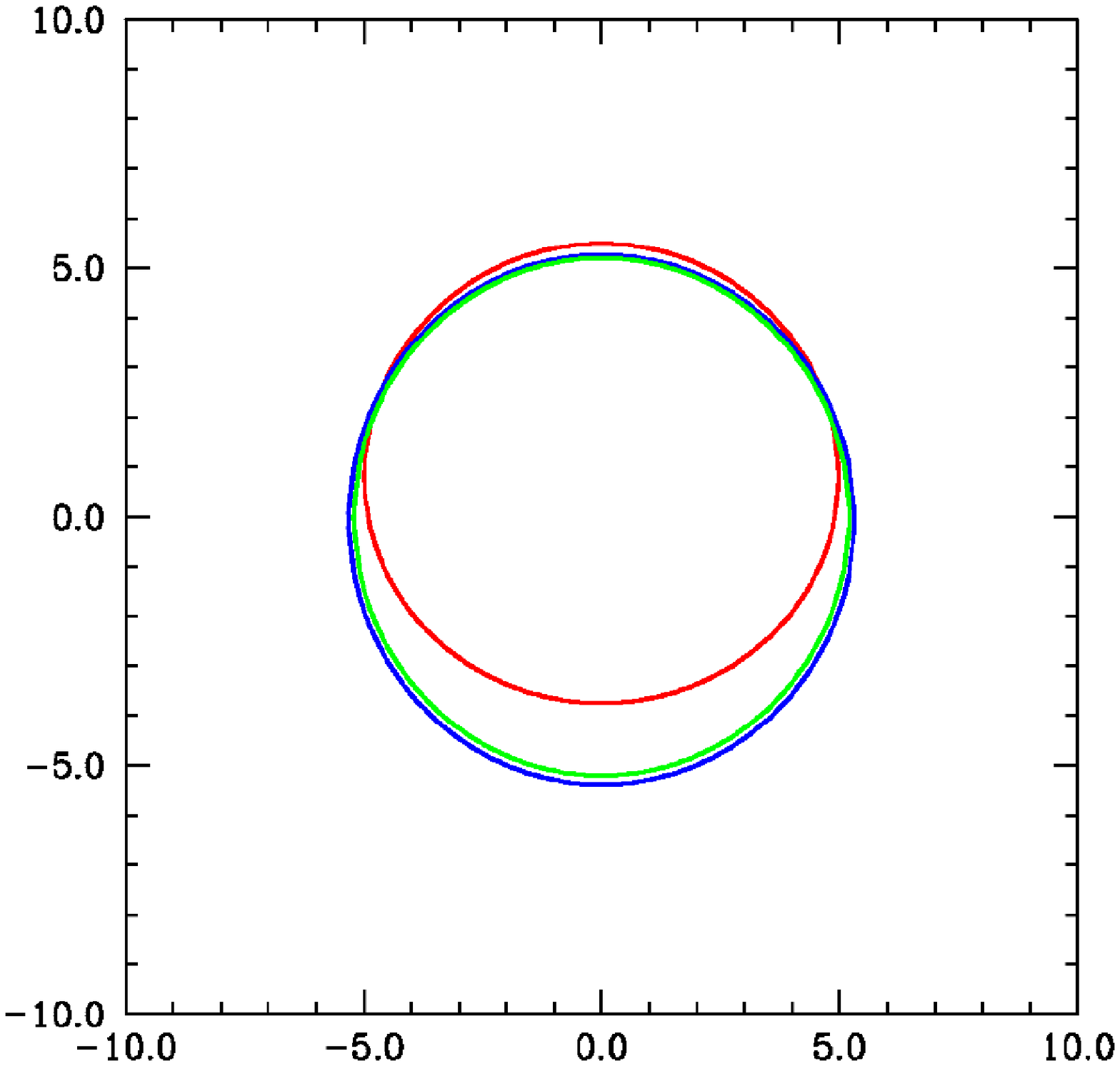}
  &
  \includegraphics[width=0.14\textwidth]{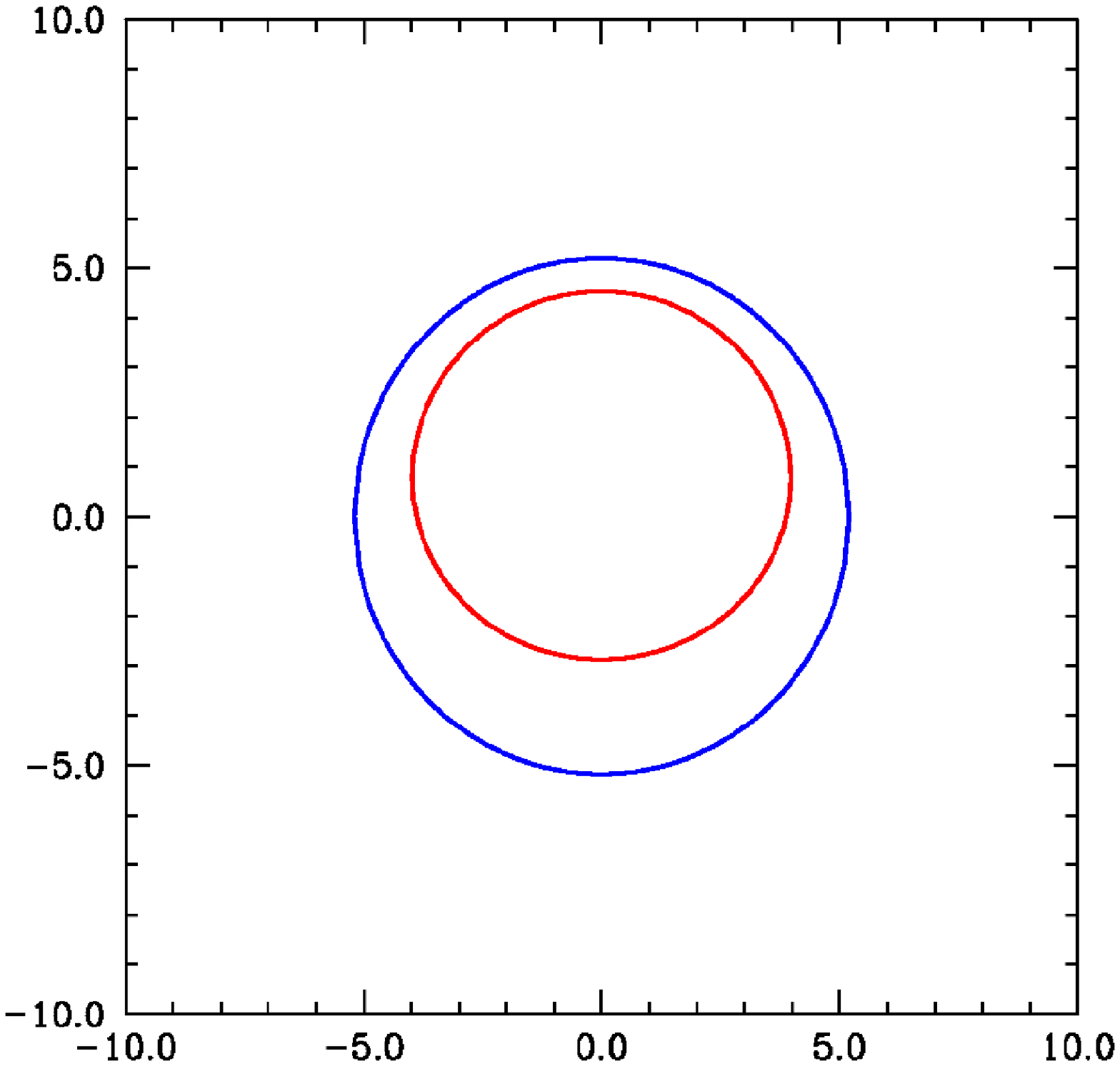}
  &
  \includegraphics[width=0.14\textwidth]{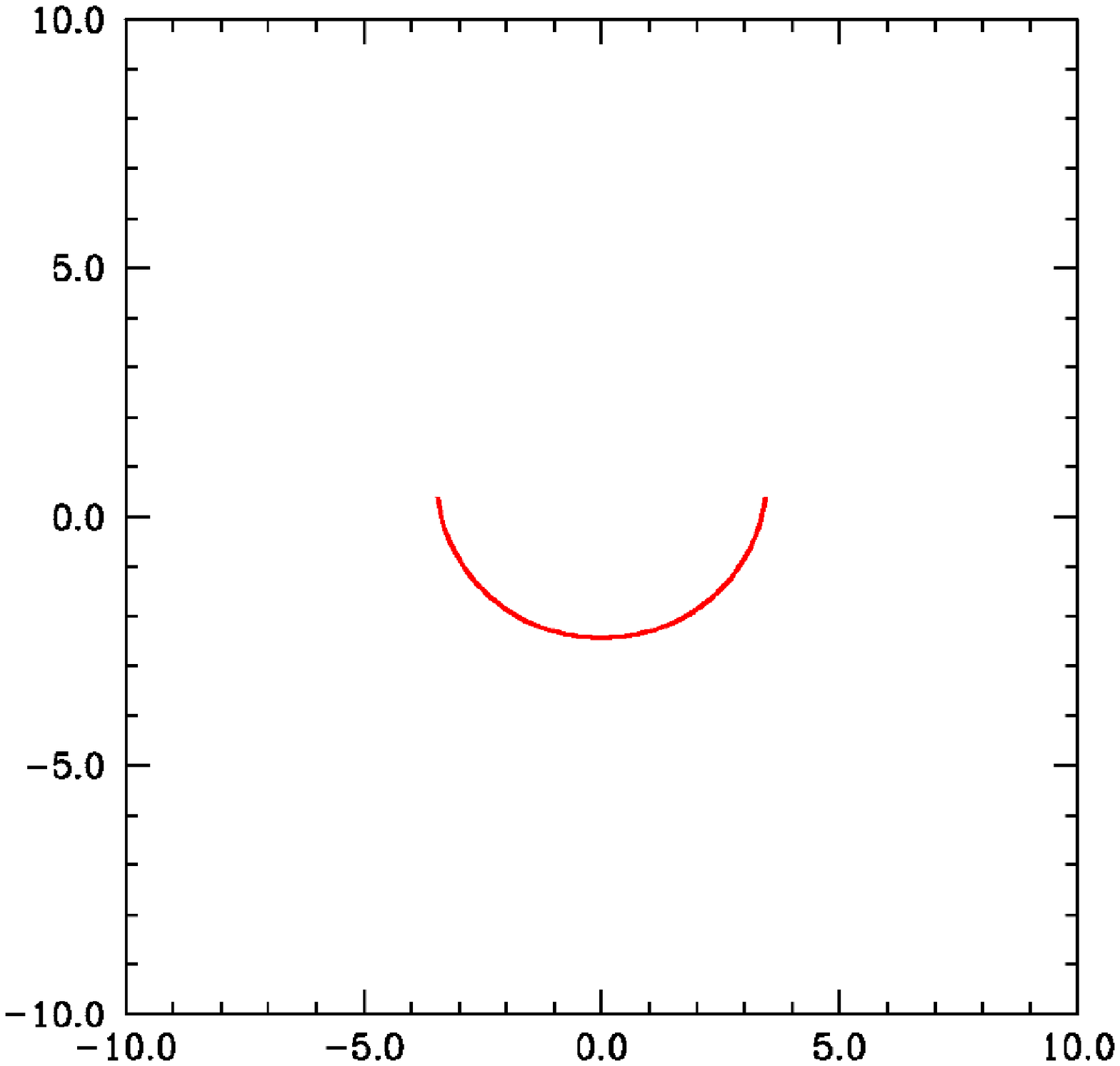}
  &
  \includegraphics[width=0.14\textwidth]{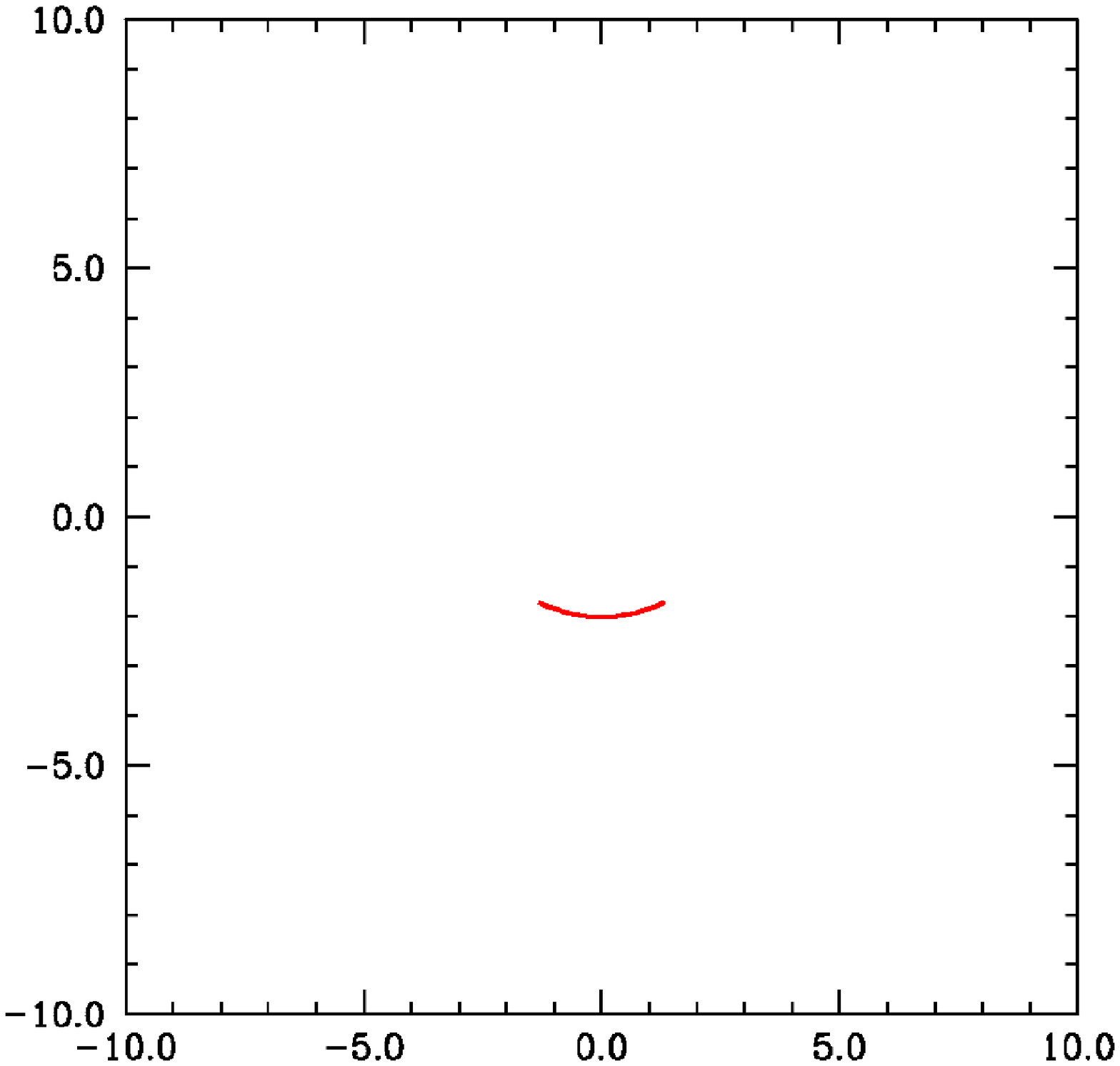}
  \end{tabular}
  \caption{As in Figure \ref{fig:3.1.1} for a Schwarzschild ($a=0$) black
hole, with (from left to right) $r_{e} = 8, 6, 4, 3, 2.5, 2 r_{g}$. The
loops display a qualitatively similar behaviour to that described in the
$a=1$ case. We note that the restoration of spherical symmetry to the
system (due to the absence of rotation) has removed the asymmetry of the
loops on both planes and hence changed the points of overlap of the loops
with respect to the $a=1$ case. For $r_{e} \le 3 r_{g}$, there are no
photons with $N=2$ that propagate to infinity, furthermore for $r_{e} < 3
r_{g}$, there are no $N=1$ photons that can propagate to infinity either.
More importantly, for $r_{e} < 3 r_{g}$ a distant observer is unable to
form a complete (that is $\phi = 0 \to 2 \pi$) image of the ring even
using direct ($N=0$) photons.}
  \label{fig:2.3.2}
  \end{center} \end{figure*}
  
\begin{figure*}
  \leavevmode
  \begin{center}
  \begin{tabular}{ccc}
  \includegraphics[width=0.3\textwidth]{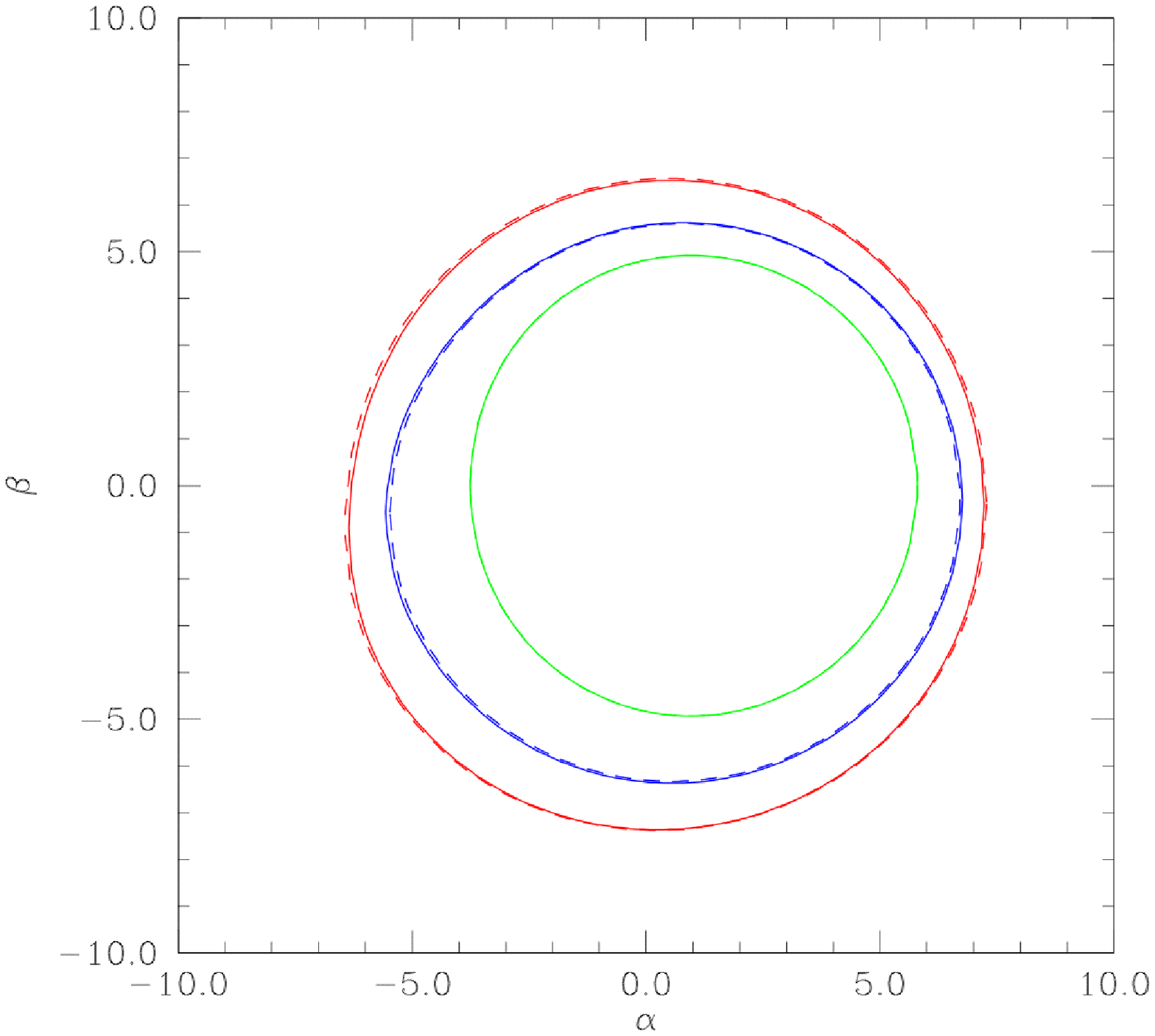}
  &
  \includegraphics[width=0.3\textwidth]{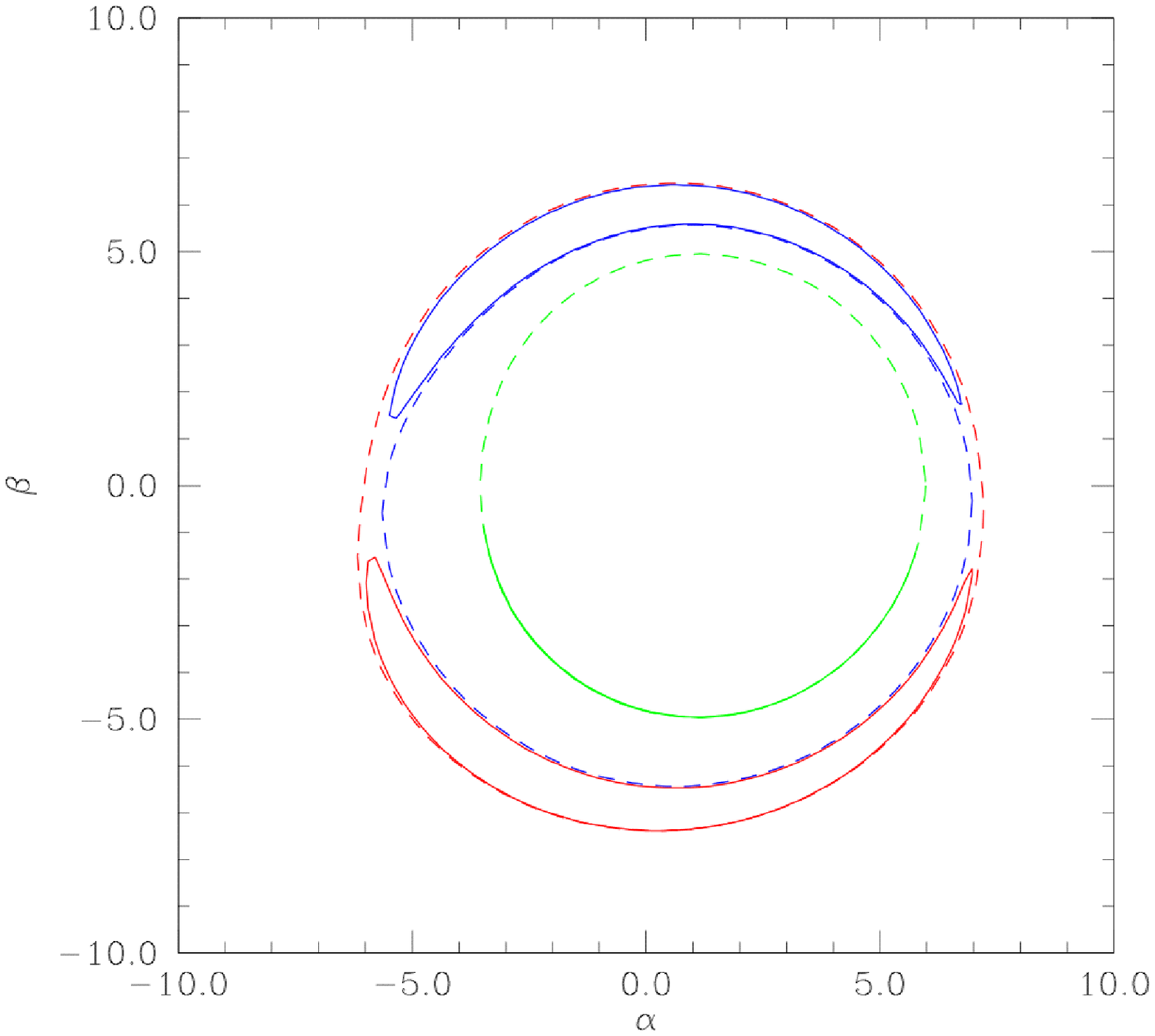}
  &
  \includegraphics[width=0.3\textwidth]{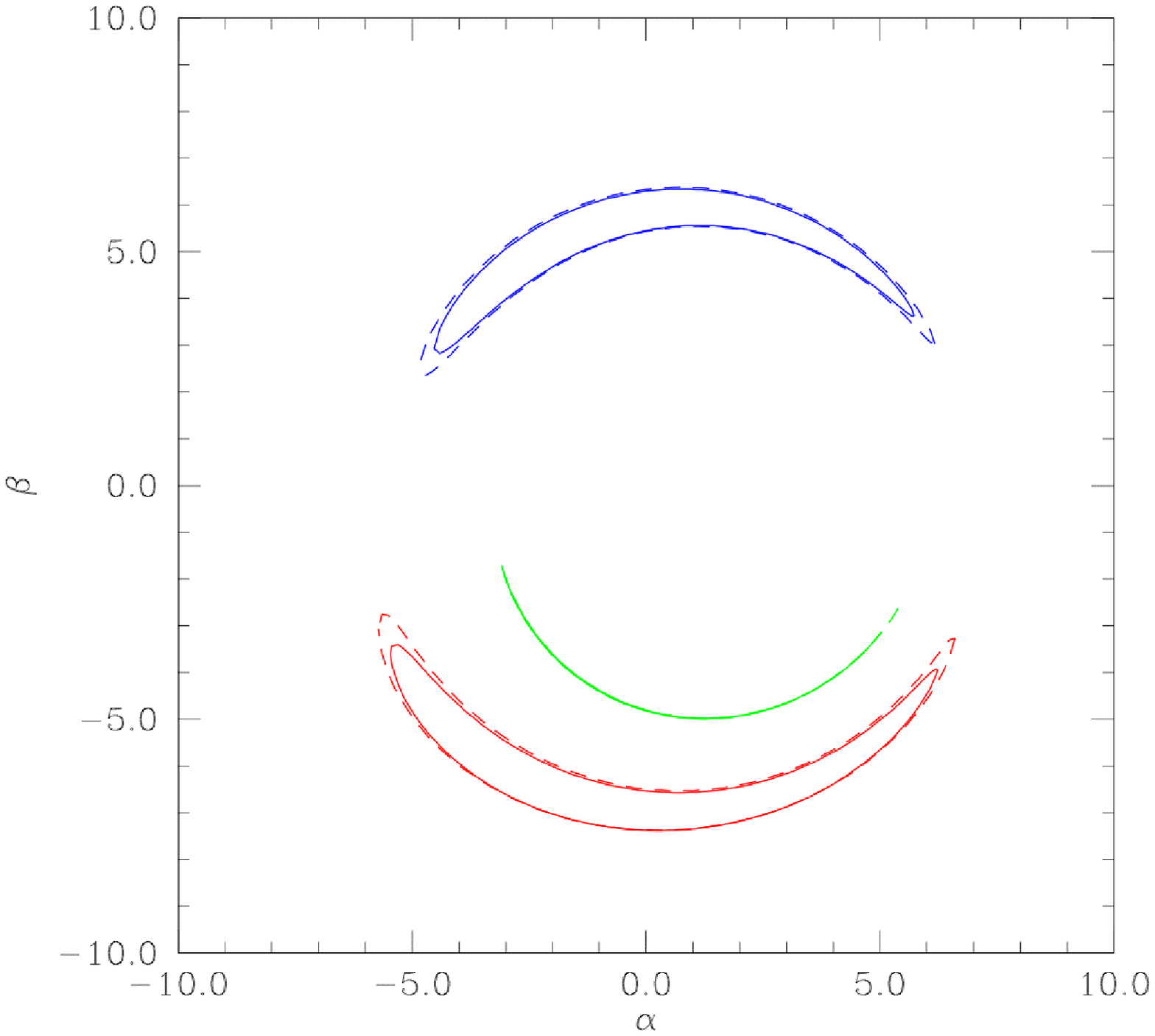}
  \end{tabular}
  \caption{Two emitting rings around an extreme Kerr black hole located at
$r_{e} = 6 r_{g}$ and $\theta^{a}_{e} = 149^{\circ}$ (solid lines),
$\theta^{b}_{e} = 151^{\circ}$ (dashed lines) with the colour coding as in
Figure \ref{fig:3.1.1}. The observer is located at radial infinity with
polar coordinate $\theta_{o} = 25^{\circ}$ (left panel), $\theta_{o} =
30^{\circ}$ (centre panel) and $\theta_{o} = 35^{\circ}$ (right panel). In
the left panel, $\left| \theta_{o} \right| < \left| \theta^{a,b}_{e}
\right|$, producing circular images of the emitting rings. In the centre
panel, we now have that $\left| \theta^{a}_{e} \right| < \left| \theta_{o}
\right| < \left| \theta^{b}_{e} \right|$ and the image of the
$\theta^{a}_{e}$ ring has now separated such that the zeroth and second
order images appear from the south pole of the hole, whilst the first
order image appears from behind the north pole. Finally, in the right-hand
panel, we have that $\left| \theta_{o} \right| > \left| \theta^{a,b}_{e}
\right|$ and so images of both $\theta^{a,b}_{e}$ follow this behaviour.}
  \label{fig:2.4.1}
  \end{center} \end{figure*}

For our purposes, three particular cases describing the behaviour of the
$q_{r,a} ( \lambda )$ curve are of interest. To illustrate these, we
consider an extreme Kerr ($a=1$) hole, with the observer located at radial
infinity and an inclination of $\theta_{o} = 30^{\circ}$. We locate the
emitter in the equatorial plane and begin by considering the case
$r^{(-)}_{ph} \le \underline{r}$ (Figure \ref{fig:2.2.1}, left panel).
Here, the region of the angular momentum $( \lambda, q)$ plane for which
no potential barrier is formed between the emitter and observer is bounded
from below by $q_{m,a} ( \lambda )$, as described by equation
\ref{eqn:2.3.4}. The upper limit of this region is provided by
$q_{\underline{r},a} ( \lambda )$ and it can be shown that, independent of
the locations of the emitter and observer, these curves are concave and
convex functions of $\lambda$, respectively. This indicates that we can
provide limits on the allowed range of $\lambda$ via: \begin{equation}
  \label{eqn:2.3.6}
  \begin{split}
    \lambda_{+,-} \left( \overline{m} , \underline{r}, a \right) = \\
    \frac{ 2 a \underline{r} \left( 1 - \overline{m}^{2} \right) \pm
\sqrt{\underline{\Delta}}
    \sqrt{1 - \overline{m}^{2}} \left( \underline{r}^{2} +
\overline{m}^{2} a^{2} \right) }
    { \underline{\Delta} - a^{2} \left( 1 - \overline{m}^{2} \right) }
  \end{split} \end{equation} The apparent angular size of the black hole
is defined by the parametric curve $q_{c} \left( r_{c} \right),
\lambda_{c} \left( r_{c} \right)$. Hence, photons whose angular momentum
falls under this curve and are directed initially inwards towards the hole
from $\underline{r}$ are inevitably captured. For $r^{(-)}_{ph} \le
\underline{r}$, this critical curve intersects the graph of $q_{m,a} (
\lambda )$ at some $\lambda^{+,-}_{c}$ (the exact intersection being
determined numerically) and so inwardly directed photons with
$\lambda^{-}_{c} \le \lambda \le \lambda^{+}_{c}$, $q_{m,a} ( \lambda) \le
q \le q_{c} ( r_{c} )$, where $r_{c}$ is determined by inversion of
$\lambda = \lambda_{c} ( r_{c})$, can be completely excluded from the
calculation.

We now move the location of the emitter inwards such that $r^{(+)}_{ph}
\le \underline{r} \le r^{(-)}_{ph} \le \overline{r}$ (Figure
\ref{fig:2.2.1}, centre panel). We have that the valid region of the
angular momentum plane is bounded from below by $q_{m,a} ( \lambda )$, as
before. The behaviour of the upper bound to the angular momentum plane is,
however, quite different. In this case, since $r^{(+)}_{ph} \le
\underline{r} \le r^{(-)}_{ph}$, there exists some $r^{*}_{c} =
\underline{r}$, which is associated with a $( \lambda^{*}_{c} , q^{*}_{c}
)$ pair, through equations (\ref{eqn:2.3.5}). For $\lambda \le
\lambda^{*}_{c}$, we therefore have that the upper limit of the angular
momentum plane is given by the critical curve and hence the lower limit on
$\lambda$ is determined by $\lambda^{+}_{c}$. Similarly, above
$\lambda^{*}_{c}$, the upper limit on the plane is given, as in the
previous case, by $q_{\underline{r},a} ( \lambda )$ and so the upper limit
on $\lambda$ is given by $\lambda_{+} ( \overline{m} , \underline{r}, a
)$. Note that, for $\lambda \le \lambda^{*}_{c}$ there are no photons that
are emitted on radially inbound geodesics that reach the observer, whilst
for $\lambda > \lambda^{*}_{c}$, only those initially ingoing photons with
values of $q$ above the critical curve can reach the observer. Finally, we
note that for the emitter located at $\underline{r} \le 2r_{g}$ (Figure
{fig:2.2.1}, right-hand panel), the graph of $q_{r,a} ( \lambda )$ is now
convex. However, the relationship between the various curves remains
unchanged from that described in the previous case and the angular
momentum plane remains bound.

\subsection{Structure \& Properties of Geodesics Loops}

We begin by considering solutions to the 'governing' equation for a simple
system where the emitter is located in the equatorial plane of an extreme
Kerr ($a=1$) hole ($\theta_{e} = \pi / 2$) and takes the form of an
infinitesimal ring located at radius $r_{e}$. For such an infinitesimal
ring, the flux is undefined (since the ring subtends zero solid angle on
the observers sky) and so we concentrate our attention on the behaviour of
the solutions on the $\left( \lambda , q \right)$ and $\left( \alpha ,
\beta \right)$ planes (Figure \ref{fig:2.3.1}). To minimise the impact of
gravitational lensing, we locate the observer at radial infinity and
$\theta_{o} = 30^{\circ}$ (as previously).

Initially, we locate the ring at $r_{e} = 8 r_{g}$ (Figure
\ref{fig:2.3.1}, panel a) and catalogue the complete set of geodesics of
zeroth ($N=0$), first ($N=1$) and second ($N=2$) order. Each of these
geodesic orders form a closed loop on both the $\left( \lambda , q
\right)$ and $\left( \alpha , \beta \right)$ planes. We note that the
projections on both planes are asymmetric about the line $\lambda, \alpha
= 0$ due to the breaking of spherical symmetry by the black hole spin and
also (for the emitter located at $8 r_{g}$) that the loops for each
individual image order are completely detached. We now move the emitter
inwards to $r_{e} = 6 r_{g}$ (Figure \ref{fig:2.3.1}, panel b), which
causes the loops associated with the zeroth and first order images to
overlap. The solutions to the geodesic equations are therefore multivalued
at these points in $\left( \lambda , q \right)$ space, breaking the
statement by \cite{C75} that there are two geodesics linking a point on
the accretion disc to an observer for each value of the redshift
parameter, $g$, which (for the Keplerian disc considered by Cunningham),
corresponds to two geodesics at each valid point in $\lambda$ space. If we
now move the emitter inwards to $r_{e} = 4 r_{g}$ (Figure \ref{fig:2.3.1},
panel c), we now see that the loops associated with the zeroth order
geodesic now overlaps with both the first order loop and the second order
loop.

We now move the emitter further inwards, so that $r^{(+)}_{ph} \le r_{e} =
3 r_{g} < r^{(-)}_{ph}$ (Figure \ref{fig:2.3.1}, panel d). Here, the
zeroth order loop still overlaps the first and second order loops,
however, in this case the first and second order loops now touch. Moving
the emitter inwards still further to $r_{e} = 2 r_{g}$ (Figure
\ref{fig:2.3.1}, panel e), the zeroth order loop now detaches itself from
the first and second order loops and moves inside these loops on the
$\left( \alpha , \beta \right)$ plane projection. The apparent angular
size of the zeroth order loop, as measured by the distant observer, is now
smaller than that of the first and second order loops, which has important
consequences for the calculation of the emergent flux, as we shall shortly
see. Finally, we move the emitter down to $r_{e} - r_{g}= 10^{-6}$ (Figure
\ref{fig:2.3.1}, panel f), the location of $r_{+}$ in Boyer-Lindquist
coordinates (however, this is not the case in terms of proper distance,
see \citealt{BPT72}). All of the loops are again detached, however, in
comparison to the case where $r_{e} = 8 r_{g}$, the ordering of these
loops is now reversed on the $\left( \alpha , \beta \right)$ plane, with
the second order now subtending the greatest angular size of the observers
sky.

We now replace the central extreme Kerr black hole with a Schwarzschild
($a=0$) black hole and repeat the preceding calculation. From Figure
\ref{fig:2.3.2}, we see that the behaviour of the solutions on the two
planes is qualitatively similar to that of the extreme Kerr case as we
move the emitter from $8 r_{g}$ down to $4 r_{g}$. However, in this case
we note that the loops are now symmetric about the line $\lambda, \alpha =
0$ due to the absence of rotation in the system and that this results in
the quantitative locations of the overlap to change. Note that we can
understand the existence of these overlaps by considering the meaning of
the projection of these loops on the $\left( \lambda , q \right)$ plane.
Recall that, for given $\left( \lambda , q \right)$ pair, there exists a
set of roots $\left( r_{i}, m_{i} \right)$ of the effective potentials
$R_{\lambda,q} \left( r \right)$, $M_{\lambda,q} \left( m \right)$.
Whether or not the geodesic path passes through selected members of these
sets of roots depends on the initial direction taken by the geodesic,
described by the $s_{r}, s_{m}$ parameters in equation \ref{eqn:2.2.4} and
hence different geodesic paths can be described by a single $\left(
\lambda , q \right)$ pair.

Consider now the behaviour of these loops as we move the emitter to $r_{e}
= 3 r_{g}$ for the Schwarzschild hole (Figure \ref{fig:2.3.2}, panel d).
Here, we find that the zeroth and first order loops are now detached, but
in this case the second order loop does not exist. Moving the emitter in
further to $2.5 r_{g}$ (Figure \ref{fig:2.3.2}, panel e), we find that not
only does the first order loop disappears, but the zeroth order loop is no
longer closed! Finally, if we move the emitter to $r_{e} - 2 r_{g} =
10^{-6}$ (Figure \ref{fig:2.3.2}, panel f), then we find that we are only
able to image a small line segment from the ring. To understand this
behaviour, we turn to the discussion given by \cite{C83} regarding the
'cone of avoidance', describing the cone generated the photons that pass
through the unstable circular orbits located at $3 r_{g}$. We let $\psi$
denote the half-angle of the cone (directed inward at large distances from
the hole) and it can then be shown that: \begin{equation}
  \label{eqn:3.1.1}
  \tan \psi = \frac{1}{ r / 3 - 1 } \sqrt{ \frac{ r / 2 - 1 } { r / 6 + 1
} } \end{equation} We therefore see that for $r \le 3 r_{g}$, $\psi \le
\pi / 2$, which implies that below $3r_{g}$, the apparent angular size of
the black hole is greater than that of the distant stars. This shows that
the black hole obscures part of the region $\phi = 0 \to 2 \pi$ for a
distant observer when the emitter is located below $r = 3 r_{g}$, which
implies that the dynamics of the accretion flow in this region are
extremely difficult to directly measure.

\subsection{General Emission Geometries}

\begin{figure*}
  \leavevmode
  \begin{center}
  \includegraphics[width=\textwidth]{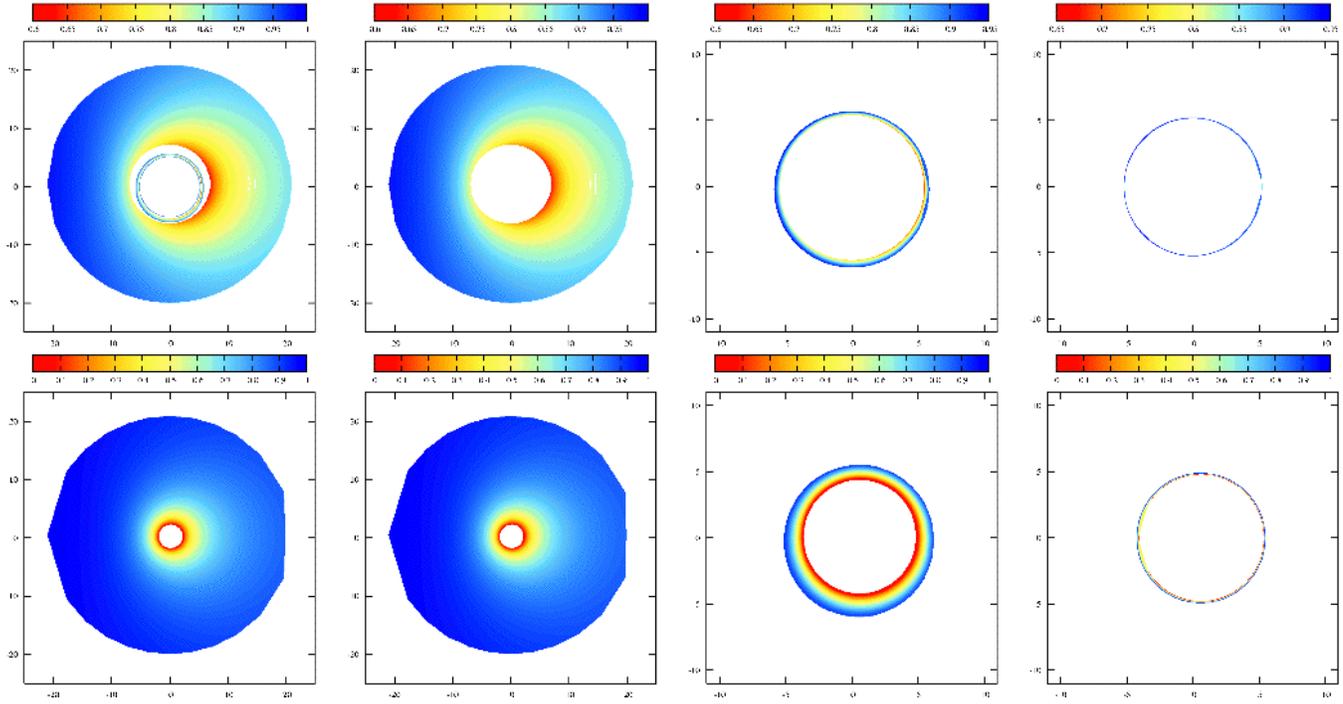}
  \caption{The contribution of orbiting photons (higher order images) to a
distant observers image of a geometrically thin, optically thick,
Keplerian accretion disc around Schwazschild (top row) and extreme Kerr
(bottom row) black holes. In both cases the observer is located at radial
infinity with $\theta_{o} = 15^{\circ}$, the disc extends from the
marginally stable orbit ($6 r_{g}$ for Schwarzschild, $1 r_{g}$ for
extreme Kerr) to $20 r_{g}$ and the images are coloured by the associated
value of the redshift parameter, $g = E_{o} / E_{e}$. From left to right,
the panels show the contributions from (a) all image orders ($N = 0 \cdots
2$), (b) the direct ($N = 0$) image, (c) the first order ($N=1$) image and
(d) the second order ($N=2$) image. Note that in the Schwarzschild case
whilst the orbiting photons are not obscured from the observer by the
accretion disc, the total flux (integrated area) measured by the observer
from the photons is negligible. Additionally, for the Kerr case, the
increased radial extent of the disc completely obscures the contribution
of the photons.}
  \label{fig:3.1.1}
  \end{center} \end{figure*}

\begin{figure*}
  \leavevmode
  \begin{center}
  \includegraphics[width=\textwidth]{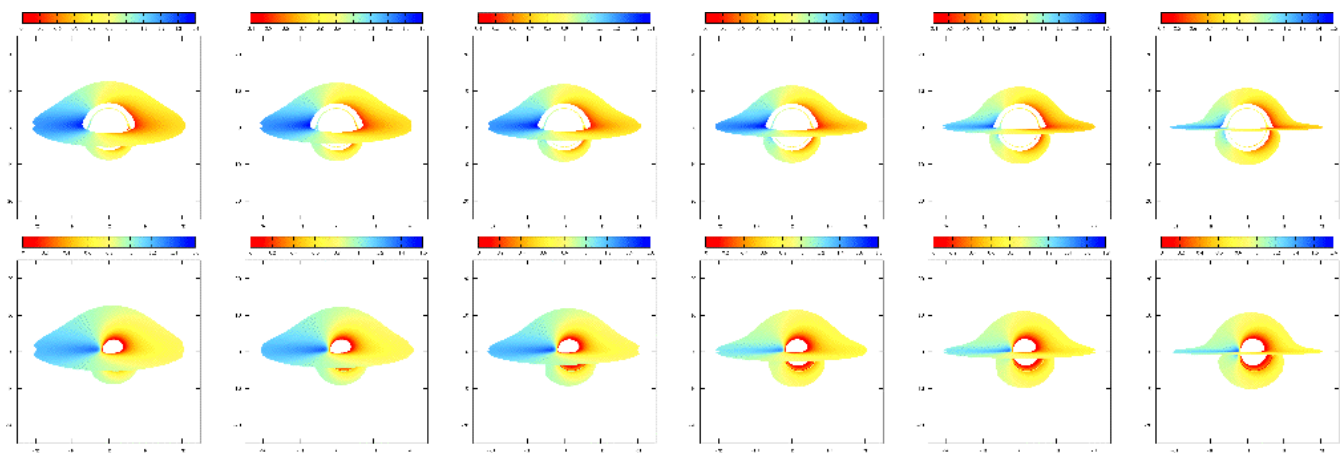}
  \caption{As in Figure \ref{fig:3.1.1} showing the contribution of
photons with $N = 0 \cdots 2$. From left to right, the panels are for
$\theta_{o} = 75.0^{\circ} \cdots 87.5^{\circ}$ in steps of $2.5^{\circ}$.
As the inclination of the system increases, the contribution from photons
emitted from the near side of the accretion disc directly to the observer
is suppressed, reducing the obscuration of the higher order images by the
disc as measured by the observer. The total area subtended by the
unobscured first order ($N=1$) image increases independently of this
effect, due to lensing of photons that are emitted from the underside of
the far side of the accretion disc, relative to the observer. In the case
of the Schwarzschild hole (top row), we note that the (total) area of the
second order image remains approximately constant with inclination. For
the extreme Kerr hole (bottom row), photons forming the second order image
are completely absorbed by the disc, removing their contribution from the
overall image received by the observer. Additionally, in contrast to the
Schwarzschild case, the images are asymmetric about the $\alpha = 0$ axis
due to the effect of black hole spin, particularly noticeable for the
higher order cases. For the most extreme inclination, the total area
subtended by the first order image is on the same order as that from the
direct image.}
  \label{fig:3.1.2}
  \end{center} \end{figure*}

In the standard picture of accretion onto a massive compact object, the
emitting material is located in the equatorial plane in what is assumed to
be a geometrically thin structure. As such, gravitational lensing effects
only come into play for high inclination systems ($\theta_{o} >
60^{\circ}$). However, if one wishes to consider emission from a
non-equatorial geometry (the case of a geometrically thick, optically thin
accretion flow, for example), then gravitational lensing effects can have
important consequences even for low inclination observers. As an example,
consider again two infinitesimal rings, located at $r_{e} = 6 r_{g}$, with
polar coordinates $\theta^{a}_{e} = 149^{\circ}$ and $\theta^{b}_{e} =
151^{\circ}$. We again locate the observer at radial infinity and
initially take $\theta_{o} = 25^{\circ}$ such that $\left| \theta_{o}
\right| < \left| \theta^{a,b}_{e} \right|$ (Figure \ref{fig:2.4.1} left
panel). The two rings form are mapped continuously onto the $\left(
\alpha, \beta \right)$ plane for each individual image order in a similar
fashion to those considered previously. We now move the latitudinal
coordinate of the observer to $\theta_{o} = 30^{\circ}$, such that $\left|
\theta^{a}_{e} \right| < \left| \theta_{o} \right| < \left| \theta^{b}_{e}
\right|$ (Figure \ref{fig:2.4.1} centre panel). The image of the lower
ring, $\theta^{b}_{e}$ is again mapped continuously onto the $\left(
\alpha, \beta \right)$ plane for each individual image order, as would be
expected as it's relationship to the latitudinal coordinate of the
observer has remained unchanged. This is not true however of the image of
the upper ring, $\theta^{a}_{e}$, which is now mapped discontinuously onto
the $\left( \alpha, \beta \right)$ plane, with the even image orders
(zeroth and second) appearing from the southern hemisphere of the hole and
the odd image orders (first) appearing from the northern hemisphere. It is
therefore clear that, in order to generate a complete picture of the
physical properties of such a system, we must include the contribution of
the higher orders to the calculation. Finally, we again move the observer
in the latitudinal direction to $\theta_{o} = 35^{\circ}$, such that
$\left| \theta_{o} \right| > \left| \theta^{a,b}_{e} \right|$ (Figure
\ref{fig:2.4.1} right panel). Now the images of both rings are mapped
discontinuously onto the $\left( \alpha, \beta \right)$ plane, which
serves to further emphasise the importance of the inclusion of the higher
order images in the calculation.

\section{Results \& Discussion}

\subsection{Images of Thin Keplerian Accretion Discs}

The contribution of higher order images to the observed flux is
dependent both on the location of the observer and the angular
momentum of the hole itself, together with the assumed geometry and
emissivity of the accretion flow. For an optically thick accretion
disc then any photons which re-intersect the disc after emission will
be either absorbed (and then re-emitted) or reflected by the material.
Figure \ref{fig:3.1.1} shows the contributions of both the direct
($N=0$)and higher order ($N=1,2$) images of a geometrically thin disc
extending from $r_{ms}$ to $20 r_{g}$, viewed at $\theta_{o} =
15^{\circ}$ for both Schwarzschild and maximal Kerr black holes. The
principle effect of black hole spin for the accretion disk dynamics is
to move the location of the marginally-stable orbit, $r_{ms}$ and
hence the location of the inner edge of the accretion disc. In the
case of the Schwarzschild hole, the inner edge of the accretion disc
is located at $6 r_{g}$, above the radius of the unstable photon
orbits ($3 r_{g}$) so higher order image photons which cross the
equatorial plane below $6 r_{g}$ are not absorbed by the disc and may
be able to freely propagate to the observer. This contrasts with the
extreme Kerr hole behaviour, where the accretion disc extends down to
$1 r_{g}$, completely obscuring the allowed radial range of the
unstable photon orbits ($1 r_{g} \le r_{c} \le 4 r_{g}$, though mostly
$3r_g$ for the polar orbits which dominate the low inclination image)
and hence the contribution of the higher order photons is completely
obscured by the accretion disc.

Figure \ref{fig:3.1.2} shows the same systems viewed at a range of large
inclination angles, $\theta_{o}= 75^{\circ}-87.5^{\circ}$.  As in the
previous discussion for the Schwarzschild hole, the inner edge of the
accretion disc is located above the location of the photon orbits and
hence the majority of the orbiting photons are able to propagate freely to
the observer. However, photons in the first order image of the far side of
the disc now have paths which pass rather closer to the hole than at lower
inclination, so the importance of lensing is increased, strongly
amplifying this part of the image. Most of these photons cross the
equatorial plane at $r \ge 20 r_{g}$, so can be seen in our simulation,
but would be obscured by a more physically realistic disc which is not
entirely flat and has outer edge $r \gg 20 r_{g}$.  These photons instead
would illuminate a large region of the underside of the disc as the direct
image of the disc, adding to its intrinsic emission.  By contrast, the
area on the sky of the second order image remains approximately constant
with increasing inclination, reflecting the sensitivity (i.e. instability)
of the two-loop photon orbits.

\begin{figure}
  \begin{center}
  \includegraphics[width=\columnwidth]{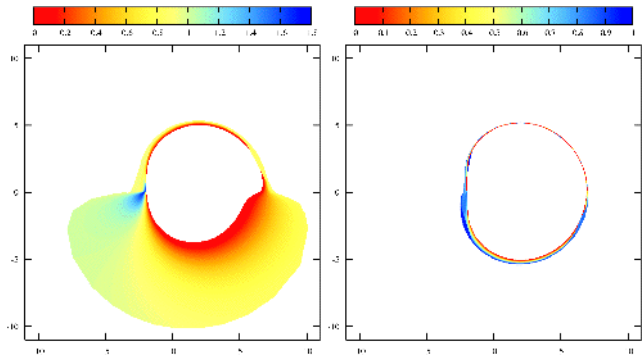}
  \caption{Unobscured $N = 1$ (left-hand panel) and $N = 2$ (right-hand
panel) images of a thin, Keplerian accretion disc around an extreme Kerr
black hole, with the disc extending from $r_{ms}$ to $20 r_{g}$. The
observer is located at radial infinity with an inclination of $\theta_{o}
= 87.5^{\circ}$. The images are highly asymmetric, indicating the strong
coupling of these photons to the bacl hole spin}
  \label{fig:3.1.3}
  \end{center} \end{figure}

The high inclination extreme Kerr images are shown in Figure
\ref{fig:3.1.2}, bottom row. The disc itself blocks all the higher orbit
images close to the black hole, similar to the $\theta_o=15^{\circ}$ case.
Part of the first order image where the geodesic crosses the equatorial
plane at $r\ge 20 r_g$ can be seen, and this fraction increases with
increasing inclination of the observer. Indeed, for the highest
inclination system considered in this work, the apparent angular size of
the first order image is approximately equivalent to that subtended by the
direct image. By contrast to the Schwarzschild case, we note that the
images of the extreme Kerr system are strongly asymmetric about the
horizontal axis due to the effect of the black hole spin, with the degree of
this asymmetry increasing with the inclination of the observer. This
asymmetry is most pronounced for the (unobscured) first and second order
images generated from the highest inclination system (see Figure
\ref{fig:3.1.3}). Here, the effect of the range of radii for orbiting
photons can be seen most clearly in the second order image. The image is
offset from zero as the ones on the left are the photons which are
emitted from the side of the disc approaching the observer, so go with
the spin of the black hole and orbit at $r_g$, while 
photons on the right are the retrograde
ones at $4r_g$. The hole in the inner disc is symmetric at $r_g$, so it
(just) obscures all the prograde photons and easily obscures the
retrograde ones. A small decrease in spin (e.g. to a=0.998, the
equilibrium spin for thin disc accretion) increases $r_{ms}$ to
$1.23$, while the photon orbits span $1.07-3.998 r_g$, so a small part of
the prograde higher order image can be seen. The fraction of higher order
photons which can escape through the 'hole' increases with increasing spin
until $a \lesssim 0.65$ \cite{BPT72}, where the photon orbits are all
within $r_{ms}$.

\subsection{Spectral Properties of Higher Order Images}

\begin{figure*}
  \leavevmode
  \begin{center}
  \includegraphics[width=\textwidth]{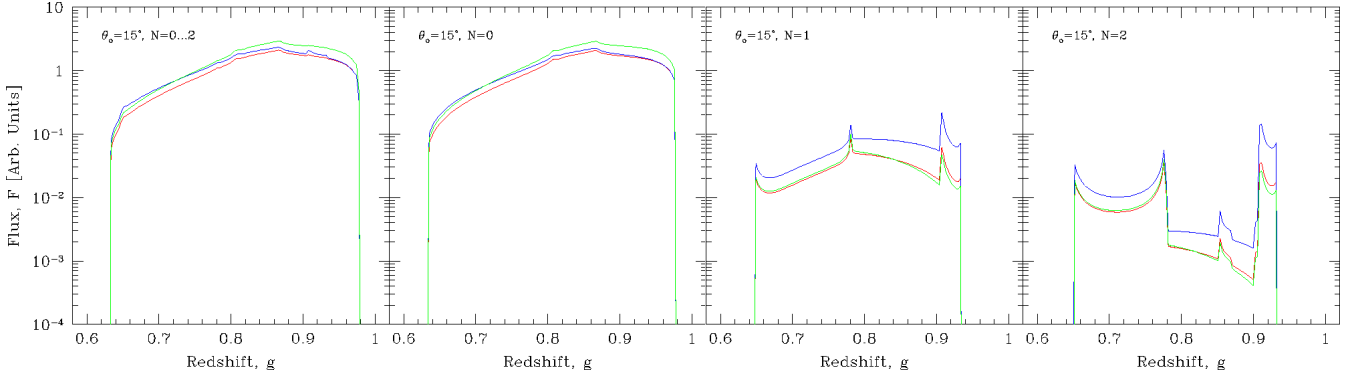}
  \caption{Relativistic line profiles generated from images (Figure
\ref{fig:3.1.1}, top row) of a geometrically thin, Keplerian accretion
disc around a Schwarzschild $(a=0)$ black hole. The observer is located at
$\theta_{o} = 15^{\circ}$ and the disc extends from $r_{ms}$ to $20
r_{g}$. All profiles are normalised such that they contain one photon and
are unsmoothed.  From left to right, the panels show the contributions
from (a) photons with $N = 0 \cdots 2$, (b) $N = 0$, (c) $N = 1$, (d) $N =
2$. In all cases a $\epsilon ( r_{e} ) \propto r^{-3}_{e}$ radial
emissivity has been applied to the line profile. We consider three
different types of angular component (i) $f ( \mu_{e} ) \propto
\mathrm{cons.}$ (red lines), (ii) $f ( \mu_{e} ) \propto \mu^{-1}_{e}$
(green lines) and (iii) $f ( \mu_{e} ) \propto ( 1 + 2.06 \mu_{e} )$ (blue
lines). We display the lines generated from the direct image (red line),
first order (blue lines), second order (green lines) and the combined line
profile (black line) using a standard $\varepsilon ( r_{e} ) \propto
r_{e}^{-3}$. the principle effect of the higher order images is to boost
the overall flux in the system by $\sim 4 \%$, whilst leaving the overall
shape of the line approximately unchanged.}
  \label{fig:3.2.1}
  \end{center} \end{figure*}

\begin{figure*}
  \leavevmode
  \begin{center}
  \includegraphics[width=\textwidth]{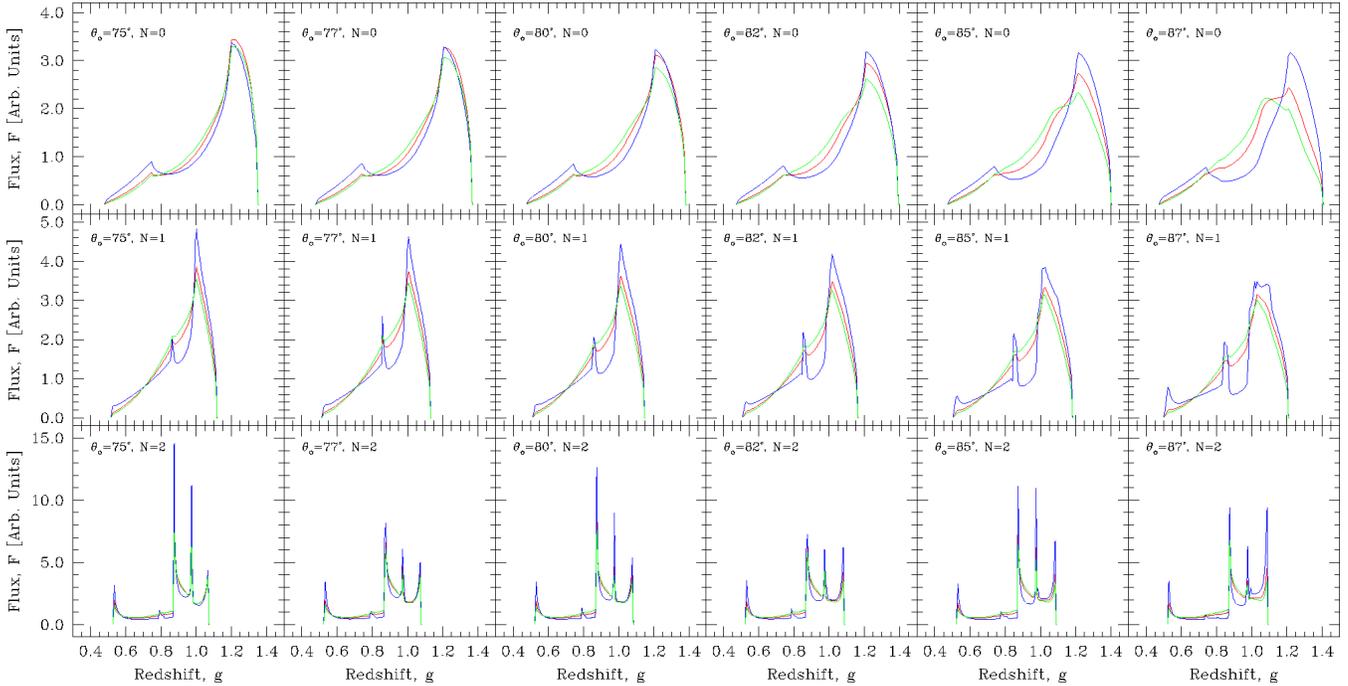}
  \caption{The contribution of orbiting photons (higher order images) to
the Iron K$\alpha$ line profiles for a geometrically thin, Keplerian
accretion disc around a Schwarzschild black hole. In all cases, the inner
edge of the accretion disc is located at the innermost stable orbit, which
occurs at $6 r_{g}$ and the outer edge of the disc is located at $20
r_{g}$. The observer is located at radial infinity with inclination (from
left to right) $\theta_{o} = 75^{\circ} \cdots 87.5^{\circ}$ in
$2.5^{\circ}$ steps. All profiles are normalised to the apparent area on
the observers sky and are unsmoothed. The line profiles are generated by
photons in (a) the zeroth order (direct) image (top row), (b) the first
order image (middle row) and (c) the second order image (bottom row).
Obscuration by the disc is ignored. In all cases a $\epsilon ( r_{e} )
\propto r^{-3}_{e}$ radial emissivity has been applied to the line
profile. We consider three different types of angular component (i) $f (
\mu_{e} ) \propto \mathrm{cons.}$ (red lines), (ii) $f ( \mu_{e} ) \propto
\mu^{-1}_{e}$ (green lines) and (iii) $f ( \mu_{e} ) \propto ( 1 + 2.06
\mu_{e} )$ (blue lines). Line profiles generated by the zeroth order
photons have the standard skewed, double peaked structure. Those generated
by the first order photons have a similar structure, whilst those from the
second order photons are far more complex. As the observer moves towards
higher inclinations, gravitational lensing effects become apparent in the
line profile generated from zeroth order photons. There is a corresponding
increase in the complexity of the profile generated from the first order
photons. Note however, that the structure of the profile generated from
the second order photons remains approximately constant with increasing
inclination.}
  \label{fig:3.2.2}
  \end{center} \end{figure*}

\begin{figure*}
  \leavevmode
  \begin{center}
  \includegraphics[width=\textwidth]{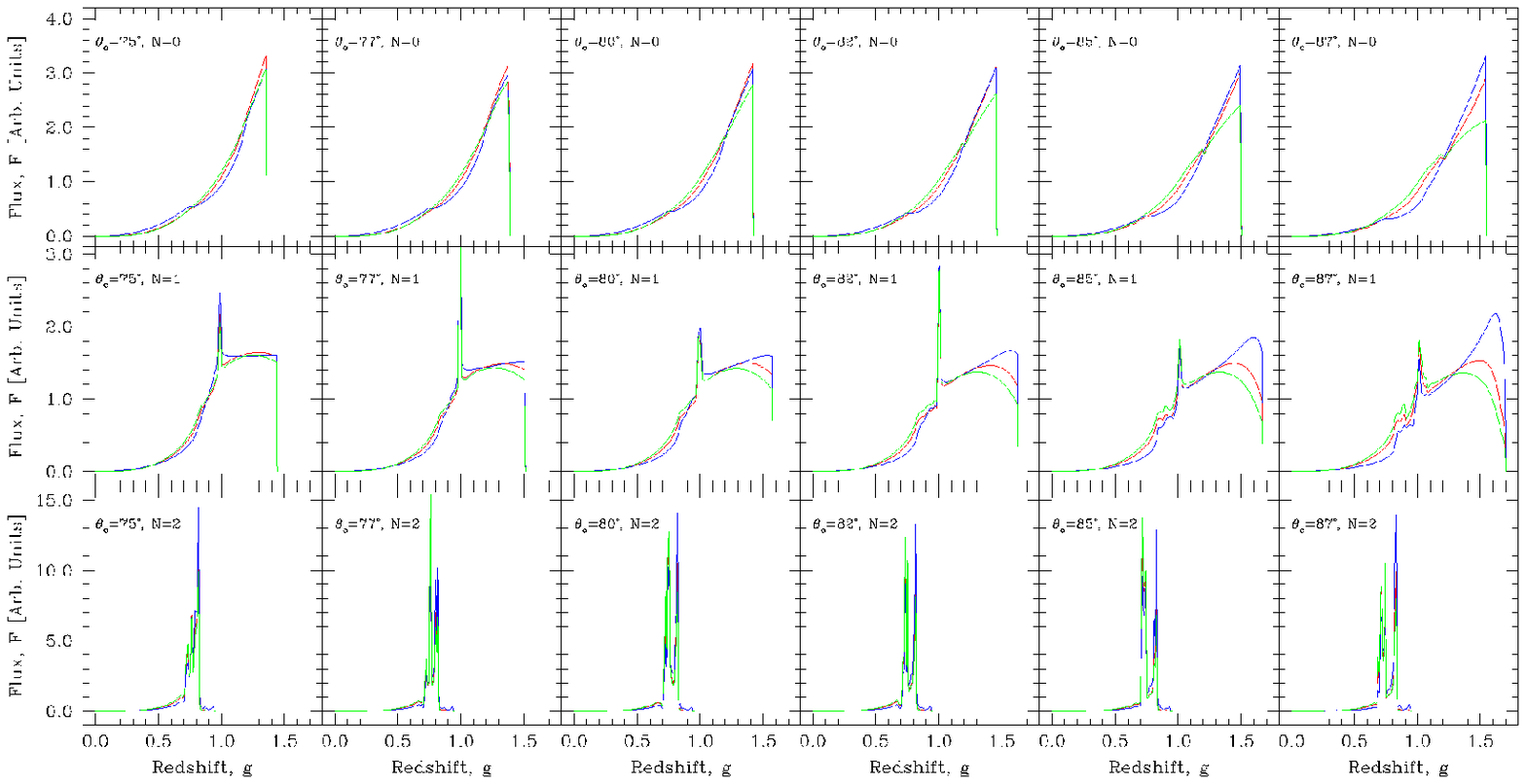}
  \caption{As in Figure \ref{fig:3.3.1} for accretion discs around an
extreme Kerr black hole. By contrast to the Schwarzschild case, the lines
extend down to lower energies due to the reduced separation between the
inner edge of the disc (at the marginally stable orbit) and the black hole
event horizon. Additionally, the line due to the first order photons
extends to higher energies than that due to the zeroth order photons, with
the upper edge of the line occuring at $g \approx 1.8$ in the most extreme
case.. Both the first and second order line profiles have significantly
increased structure compared to the zeroth order profile. The shape of the
zeroth and second order line profiles remain approximately constant with
increasing inclination. However, the internal structure of the first order
profile shows a distinctive change, with the fraction of radiation in the
blue wing of the profile exhibiting a marked increase compared to that in
the red wing as the observer moves towards the equatorial plane.}
  \label{fig:3.2.3}
  \end{center} \end{figure*}
  
To understand how the astrophysical properties of the accretion flow
couple to the gravitational field of the black hole, we generate
relativstically smeared line profiles as described in \cite{BD04}. For
clarity, we briefly recap this method here. We consider an intrisically
narrow emission line with rest energy, $E_{e}$, for which the flux
distribution measured by a distant observer at an energy $E_{o}$ is given
by: \begin{align}
  \label{eqn:3.2.1}
    F_{o} \left( E_{o} \right) = \frac{1}{D_{o}^{2}} \int \int g^{4}
\varepsilon \left( r_{e},\mu_{e} \right) \delta \left( E_{o} - gE_{int}
\right)  d\alpha d\beta \end{align} Here, $\varepsilon \left(
r_{e},\mu_{e} \right)$ is the local emissivity of the accretion disc,
which we take to have the form $\varepsilon \left( r_{e},\mu_{e} \right) =
\epsilon \left( r_{e} \right) f \left( \mu_{e} \right)$, where $\mu_{e}$
describes the initial direction of the photon relative to the local z-axis
of the emitting materlal. The flux at each redshift in each image order is
then calculated directly from the area subtended on the observers sky,
together with the intrinsic disc emissivity (both radial and angular).
This gives the transfer function for monochromatic flux, so the observed
emission is the convolution of this with the intrinsic disc spectrum. In
the case of intrinsically monochromatic radiation, e.g. the iron K$\alpha$
fluorescence line which can be produced by X-ray illumination of cool,
optically thick gas, then this transfer function directly gives the
expected line profile. Photons produced by this emission process in
regions close to the black hole enable us to examine the properties of the
multiple orbit photons considered in the preceding section from a
spectroscopic perspective.

To generate the line profiles, we apply a radial emissivity of the form
$\epsilon ( r_{e} ) \propto r^{-3}_{e}$ (consistent with gravitational
energy release within the disc, \citealt{ZDS98}) and consider three
possible angular emissivity laws: (i) $f ( \mu_{e} ) \propto
\mathrm{cons.}$, corresponding to an optically thick disc (red lines);
(ii) $f ( \mu_{e} ) \propto \mu^{-1}_{e}$, corresponding to an optically
thin, limb brightened disc (blue lines) \citep{MPS93} and (iii) $f (
\mu_{e} ) \propto ( 1 + 2.06 \mu_{e} )$ corresponding to an optically
thick, limb darkened disc (green lines) \citep{LNP90}. Figure
\ref{fig:3.2.1} shows the line profiles generated for the low inclination
Schwarzschild disc discussed in the preceding section. Whilst in the case
of the extreme Kerr black hole (Figure \ref{fig:3.1.1}, bottom row), all
orbiting photons returned to the inner region of the disc, for the
Schwarzschild hole (Figure \ref{fig:3.1.1}, top row) these photons cross
the equatorial plane below $r_{ms}$ and can freely propagate to the
observer (although see \citealt{RB97}). Examination of their spectral
properties shows that these photons play a limited role in forming the
overall spectral shape measured by a distant observer, which is completely
dominated by the contribution from the direct image.

Figure \ref{fig:3.2.2} shows the line profiles generated for a
Schwarzschild hole, with the accretion disc now viewed at high inclination
(ignoring the effect of obscuration). For the direct image, limb darkening
boosts the effects of gravitational lensing, enhancing the flux from the
far side of the hole. This is because these photons are strongly bent,
i.e. are emitted from a lower inclination angle than that at which they
are observed, so a limb darkening law means that the flux here is higher
\citep{BD04}. The doppler shifts are rather small for this material, so
this lensing enhances the flux in the middle of the line. Since the line
profiles are normalised to unity, this means that the blue wing is less
dominant.

The first order spectra are shown in Figure \ref{fig:3.2.2}, middle row.
The transfer functions mostly retain the characteristic double peaked and
skewed shape, and again the principle effect of the different angular
emissivities is to alter the balance between the blue wing and lensed
middle of the line.  However, there is some new behaviour for the limb
brightened emissivity.  This has the largest change in emissivity with
angle, and this combined with the exquisite sensitivity of lensed paths
means that this picks out only a small area on the disc, leading to a
discrete feature in the spectrum. These profiles also show enhancement of
the extreme red wing of the line, as the photons which orbit generally are
emitted from the very innermost radii of the disc.

The discrete features are completely dominant for all emissivities at
second order (Figure \ref{fig:3.2.2}, bottom row). These are images of the
top of the disc where the photons have orbited the black hole, so the
paths are even more sensitive to small changes than first order. Thus the
profiles are significantly more complex in structure, being dominanted by
lensing.  There are blue and red features at the extreme ends of the line
profile which are picking out the maximum projected velocity of the
innermost radii of the disc. These have the standard blue peak
enhancement. However, the two strong features redward of this are a pair
of lensed features, from the near and far side of the disc.

Figure \ref{fig:3.2.3} shows the line profiles generated for the extreme
Kerr hole at high inclinations, again ignoring the effect of obscuration.
For the direct image (Figure \ref{fig:3.2.3}, top row), the lines exhibit
the characteristic triangular shape previously reported by (e.g)
\cite{L91}, with the variation in angular emissivity acting to alter the
balance between the different regions of the line on a $\sim 5 \%$ level.
The lines associated with the first order image (Figure \ref{fig:3.2.3},
middle row) exhibit a marked difference in comparison to those associated
with the Schwarzschild black hole. In general, they are broader that those
associated with the direct image and resemble a skewed Gaussian combined
with a narrow line (due to caustic formation) at $g \approx 1.0$. Here the
principle effect of changes in the angular emissivity is to alter the
height of the blue wing, relative to the rest of the line. Again, the line
profiles associated with the second order image (Figure \ref{fig:3.2.3},
bottom row) are completely dominated by discrete features, as in the
Schwarzschild case.

\subsection{Image Luminosities}

\begin{figure}
  \begin{center}
  \includegraphics[width=\columnwidth]{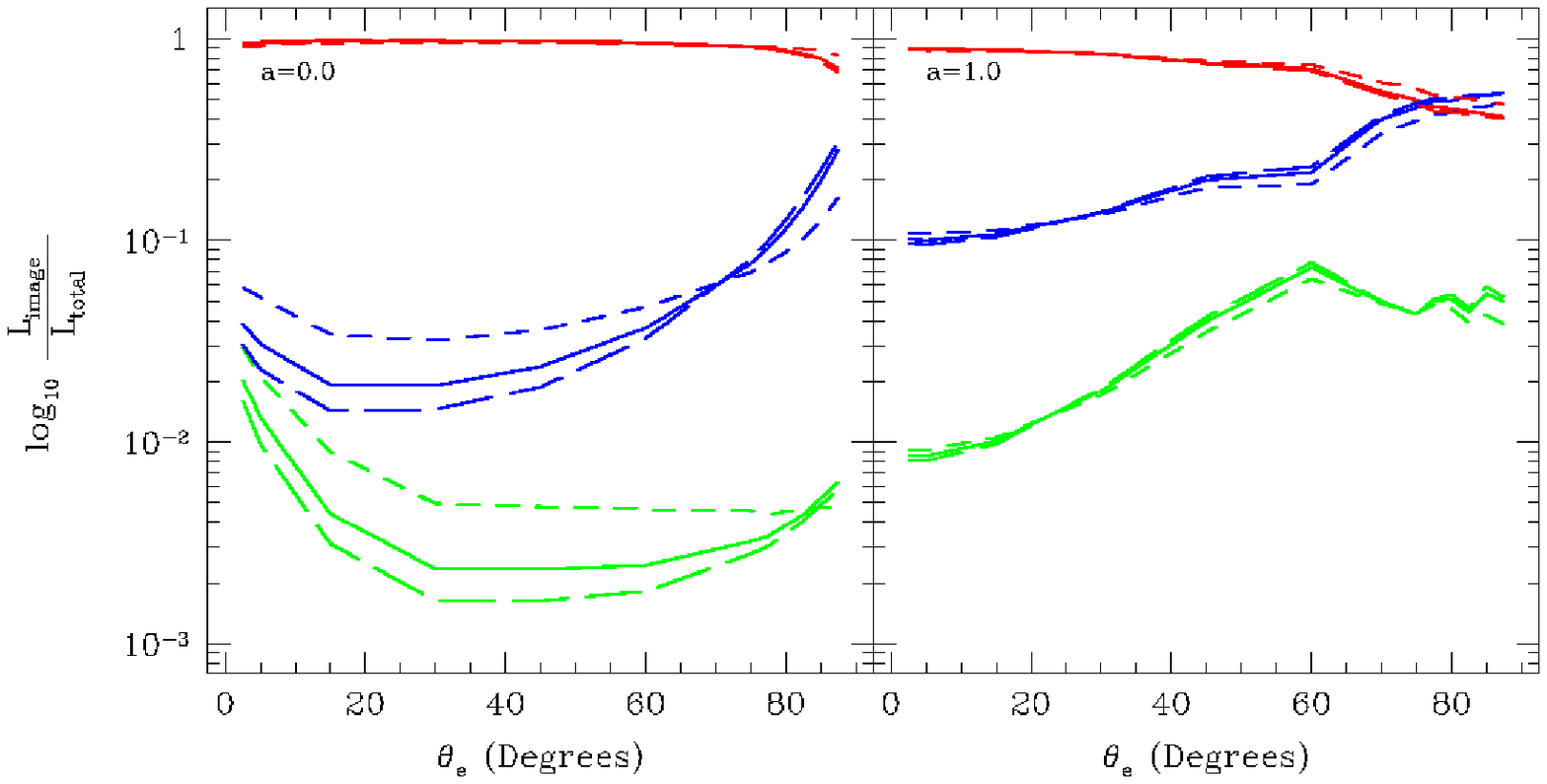}
  \caption{Variation of (unobscured, integrated) luminosity associated
with each image as a fraction of the total luminosity of the system with
inclination for a thin, Keplerian accretion disc around a Schwarzschild
black hole (left panel) and extreme Kerr (right panel). In both cases, the
disc extends from $r_{ms}$ to $20 r_{g}$ and the observer is located at
radial infinity. We show the contributions from (a) the direct image (red
lines), (b) the first order image (blue lines) and (c) the second order
image (green lines). In all cases, we have applied a $\epsilon ( r_{e} )
\propto r^{-3}_{e}$ radial emissivity and have considered angular
emissivities of the form (i) $f ( \mu_{e} ) \propto \mathrm{cons.}$ (solid
lines), (ii) $f ( \mu_{e} ) \propto \mu^{-1}_{e}$ (short dashed lines) and
(iii) $f ( \mu_{e} ) \propto ( 1 + 2.06 \mu_{e} )$ (long dashed lines). In
the case of the Schwarzschild hole, the contribution from the higher order
images is $\le 20 \%$ of the total luminosity, even at high inclinations.
However, in the case of the extreme Kerr black hole, at inclinations of
$\ge 75^{\circ}$, the first order image can dominate the luminosity of the
system, due to the formation of caustics within the image.}
  \label{fig:3.3.1}
  \end{center} \end{figure}

To understand the relative roles played by each (unobscured) image order,
we consider the variation of the luminosity of each image as a fraction of
the total luminosity of the system as a function of inclination, again for
both Schwarzschild and extreme Kerr black holes (Figure \ref{fig:3.3.1}).  
These luminosities are generated from the integral in redshift space of
the line profiles considered in the preceding section and hence we
consider systems with identical properties to those previously discussed.
In the case of the Schwarzschild hole, we see that, for inclinations $
<80^{\circ}$, the first order image can be regarded as a $\le 10 \%$
correction to the emergent flux from the system. For inclinations $\ge
80^{\circ}$, this image contains $~ 10 - 20 \%$ of the emergent flux, i.e.  
even at these high inclinations, the first order image can still be
regarded as a correction to the direct image. For the second order image,
we see that it plays a $\le 1 \%$ role independent of inclination.  Third
and higher order images will have correspondingly smaller fluxes, so can
safely be neglected.

There is a larger fraction of flux in the higher order images for the Kerr
black hole. At low inclinations, the first and second order images contain
$~ 10 \%$ and $~ 1 \%$ of the total flux, respectively. The variation of
the luminosity fraction for the three image orders displays an
approximately power law like behaviour for inclinations $\le 60^{\circ} $,
where a distinct break occurs, due to the appearance of a caustic in the
first order image, whose luminosity is therefore significantly enhanced.
It is at this point that the peak luminosity of the second order image
occurs, which here is on the level of $~ 10 \%$, approximately an order of
magnitude higher than in the Schwarzschild case. Remarkably, for
inclinations $> 75^{\circ}$, the luminosity of the first order image
produced by the optically thick discs is \emph{greater} than that produced
by the direct image.

However, most of the higher image order flux is expected to reintercept
the disc plane, and hence be absorbed and reradiated. This is especially
the case for realistic discs around an extreme Kerr black hole, where the
whole of the equatorial plane is covered by the disc from $r_g$ to large
radii. However, for Schwarzchild, the existance of a central 'hole' means
that the flux from higher order images can escape. A realistic disc around
a Scharwzchild black hole when viewed {\em face on} has $~ 10 \%$ of its
flux in a higher order image ring (dropping to $\sim 7\%$ for a more
extended disk ranging from $r_{ms} \to 400 r_{g}$). Essentially, a spatial
resolution equivalent to $2r_{g}$ is required to resolve these features,
which for typical nearby Active Galactic Nuclei corresponds to an angular
resolution of $~ 0.01$ microarcseconds. This is obviously extremely
technically difficult, but is feasible for an X-ray interferometer imaging
supermassive black holes in nearby galaxies \citep{G01}.

\section{Conclusions}

Photons orbiting a black hole encounter the strongest possible
gravitational fields compatible with still escaping to the observer.
Thus these higher order null geodesics provide the best test of Einsteins
gravity in the strong field limit.  We have developed a new strong gravity
code capable of describing these paths, and calculate them for a
geometrically thin, optically thick (standard) disc in both Schwarzchild
and Kerr metrics. These higher order image paths must cross the equatorial
plane, so are absorbed where this is filled by the optically thick disc.
As has long been known, the major amplification effects of gravitational
lensing are for the first order paths from the far side of the underneath
of the disc viewed at high inclination i.e. photons initially emitted
downwards on the far side of the black hole, which are bent by gravity up
above the disc plane. Most of these paths will re-intersect the disc
unless it has very limited outer radial extent. While such discs {\em may}
exist \citep{R04}, it seems far more likely that these photons would be
absorbed by material in the equatorial plane. However, there is some
fraction of the higher order images where the light paths are so strongly
bent that they re-intersect the equatorial plane very close to the photon
orbit radius. By definition, this is below the minimum stable orbit for
particles, so the standard disc cannot exist at this point. Instead, for a
stress-free inner boundary condition, the disc material plunges rapidly
though this region, so there is much less absorbing material in the
equatorial plane. This material may be optically thick at high mass
accretion rates \citep{RB97}, but this depends on whether the flow in the
plunging region in smooth or clumpy. Hence
higher order photons which cross the plane below $6r_g$ need not
be reabsorbed by the disc. We show that these {\em observable} higher
order photons can carry 10\% of the flux for a {\em face on} disc. Edge on
discs {\em reduce} the expected observable flux as only about half of the 
orbiting photon ring can be seen through the gap below $r_{ms}$, while
the rest re-intercepts the disc at $r \gg r_{ms}$.

The situation is less favourable in the Kerr geometry as 
photons now orbit at a range of radii depending on how their
angular momemtum is aligned with the spin of the black hole. Photons going
against the spin will orbit at radii which are larger than that of the
minimum stable orbit of the disc, so should be absorbed rather than
escaping through the inner 'hole' in the disc (provided $a \gtrsim 0.65$).
Thus for Kerr black holes, even though there is a greater fraction of
photons in the higher order images, we expect that these are less
observable due to the overlap between the photon and particle orbits.
Also, even though there is still a gap between the last stable particle
orbit and the aligned photon orbits, this gap is much smaller than in
Schwarzchild, both in terms of radial coordinate and in terms of impact
parameter on the sky. For a=0.998, the equilibrium spin for thin disc
accretion, $r_{ms}=1.23 r_g$ while the $r_{ph}^{(-)}=1.07$, so $\Delta
r=0.16$.  Since the photons orbit in slightly stronger gravity, their
paths are more distorted, so the difference in their impact parameter at
infinity is slightly less than the difference in radial coordinate. Thus
the higher order images round a rapidly spinning black hole are much more
difficult to spatially resolve from the direct image of the disc, and
carry less flux than Schwarzchild as only a fraction re-intersect the
plane below $r_{ms}$.

Thus the best chance of observing these higher order photon paths from a
standard accretion disc is with a face-on Schwarzchild black hole. X-ray
interferometry could potentially directly resolve such scales for
supermassive black holes in nearby galaxies, and such missions are being
seriously proposed (MAXIM: \citealt{G01}). Even if the plunging region is
optically thick, direct imaging with resolution $\le 2r_g$ will clearly
show the black hole spin from the apparent size of the 'hole' in the
center of the disc. The radial coordinate of the innermost stable orbit in
an equilibrium spin Kerr metric is a factor of $\sim 3$ smaller than for
Schwarzchild, which translates to a change in apparent size at infinity of
the 'hole' diameter from $\sim 5 r_g$ (a=0.998) to $\sim 14r_g$ (a=0).
This is important as recent papers have emphasised that the true size
of the shadow of the event horizon (in effect the impact parameter of the
orbiting photons, given fairly accurately by our second order image) is
rather similar in both Schwarzchild and Kerr \citep{FMA00,F03,T04}. This
is true, but different to the size of the 'hole' defined by the innermost
stable orbit of the disc. For continuous energy release down to the event
horizon, the 'hole' in the image of the accretion flow is set by the true
shadow size, but for accretion flows where the energy is only released
from stable particle orbits (the disc, as opposed to the plunging region)
then the innermost stable orbit sets the size scale. Whether the energy
release can be continuous across $r_{ms}$ is currently a matter of active
research \citep{KH02}. However, there is clear observational evidence for
an innermost stable orbit in the disc dominanted spectra of galactic black
hole binaries \citep{E93,K99,GD04}, so it seems likely that the accretion flow can
take the standard thin disc form assumed here.

The spectral signatures of these photons only become apparent for
high-inclination systems, where they carry between $\sim 20 \%$
(Schwarzschild) and $\sim 60 \%$ (extreme Kerr) of the total luminosity of
the system. However, in a realistic system, the disc extends out to many
thousands of gravitational radii from the black hole and so these photons
return to the disc before reaching the observer. Far from the central
black hole, the effect of the returning radiation is comparable 
to the gravitational potential energy of the material and so these photons
can play an important role in shaping the properties of the disc here. In
particular, reprocessing of this returning radiation at large distances
from the hole will potentially provide a physical mechanism to flatten (at
large radii) the $r^{-3/4}$ temperature profile of a planar accretion disc
irradiated by a central source, which could help to resolve the 
conflict between the predicted and observed optical/UV colours in Active
Galactic Nuclei \citep{KB99}. 

The code used to calculate these results is an extension of that of
\cite{BD04}. The new aspect described here is a set of analytic
constraints on the possible photon trajectories which vastly reduce the
scale of the calculation. This, combined with the use of the analytic
integrals (elliptic functions) for the photon paths, has the result that
the calculation of a photon trajectory linking two points (the emitter and
observer, \citealt{V93}) can be performed by a simple minimisation,
capable of being calulated to almost arbitrary accuracy on a standard
desktop pc on time scales of a few minutes. This technique allows us to
include the contribution of orbiting photons in the calculations without
loss of resolution. The code is also flexible enough to be adapted to {\em
any} accretion flow, and a future paper will consider the contribution of
the higher order images in an optically thin, geometrically thick flow.

\label{lastpage}

\end{document}